# HOW GENERATIVE AI ADOPTION ALTERS THE DEMAND FOR COGNITIVE AND SOCIAL SKILLS WITHIN ROLES: A SKILL-CENTRIC ANALYSIS

Piyush Gulati
p.gulati@ucl.ac.uk
UCL School of Management

Arianna Marchetti
amarchetti@smu.edu.sg
Lee Kong Chian School of Business, Singapore Management University

Victoria Sevcenko
victoria.sevcenko@insead.edu
INSEAD

Phanish Puranam
phanish.puranam@insead.edu
INSEAD

## RESEARCH ABSTRACT

A common view holds that generative AI (GenAI) automates cognitive tasks, reshaping roles to emphasize social skills over cognitive ones. Drawing on the framework we develop in this paper, we argue that other outcomes are theoretically possible. We analyze seven million job postings from 595 U.S. public firms that adopted GenAI in 2022-2024, estimating difference-in-differences models comparing GenAI-adopting and non-adopting roles around ChatGPT's launch. We find no evidence of greater emphasis on social skills in GenAI-adopting roles; instead, their relative demand for social skills declined by 3.4%, while demand for cognitive skills held steady. This decline reflects the addition of functional administrative skills to roles outside those functions. Concentrated in non-managerial roles, this pattern points toward GenAI-enabled self-service rather than intensified cross-functional coordination.

## MANAGERIAL ABSTRACT

Many expect generative AI (GenAI) to take over analytical work, leaving people to focus on collaboration, teamwork, and other social skills. We studied seven million job ads posted by 595 U.S. public companies before and after ChatGPT's launch. Jobs that began requiring GenAI did not ask for more social skills. Instead, these jobs grew broader by adding finance, human resources, and legal skills, even in roles outside those departments, and mostly among non-managers. As a result, relative demand for social skills fell 3.4%, while relative demand for cognitive skills held steady. We interpret these results as GenAI letting employees handle administrative work themselves rather than coordinate with specialists. Companies and training programs betting on social skills as protection against automation should therefore reconsider that assumption.

# 1. INTRODUCTION

Unlike prior technologies, generative artificial intelligence (GenAI) exhibits competencies at non-routine tasks previously thought to be uniquely human (Acemoglu & Restrepo, 2019; Argote et al., 2021). This development challenges long-held assumptions about how technology may change the demand for human skills. In this paper, we specifically examine how GenAI adoption alters roles by changing their requirements for workers' cognitive and social skills.

Cognitive skills, such as analytical reasoning and problem solving, and social skills, such as coordination and negotiation, are central to workplace success (Acemoglu, 2002; Deming, 2017). These domain-agnostic skills function as "meta skills" that enable individuals to learn firm-specific routines and collaborate effectively within the organization (Anderson & Krathwohl, 2001; Becker, 1962; Wang & Barney, 2006), and the wage returns to both types of skills have grown steadily over recent decades (Deming, 2017; Weinberger, 2014). Workforce training and development policies also increasingly emphasize social and interpersonal competencies as protection against automation (Madgavkar et al., 2024; WEF, 2025). If GenAI adoption alters roles in ways that make these skills less relevant, this could diminish workers' opportunities to develop them, affect their wage returns over time, and call into question the prevailing policy emphasis on social skills as the durable complement to technology.

One widely held view is that because GenAI's unique capabilities lie largely in cognitive tasks, its adoption will automate work that requires cognitive skills, while allowing employees to focus on work that requires social skills (Hernholm, 2025; PwC, 2026; Raman & Flynn, 2024). For instance, Madgavkar et al. (2024) project that demand for social and emotional skills will rise by 11-14 percent by 2030 and demand for roles relying on basic cognitive skills will decline by a similar margin. However, in this paper, we argue that several other trajectories are also plausible.

Prior work on the labor market effects of GenAI examines how this technology impacts worker productivity (e.g., Brynjolfsson et al., 2025; Dell'Acqua et al., 2023; Hoffmann et al., 2024; Noy & Zhang, 2023), wages (e.g., Lichtinger & Hosseini Maasoum, 2025), and labor demand (e.g., Humlum & Vestergaard, 2025; Lichtinger & Hosseini Maasoum, 2025). However, these studies do not directly explain how the technology reshapes roles in terms of their skill requirements. A role is a bundle of tasks, and the individuals occupying it bring their skills to bear on those tasks. A role's skill requirements may remain unchanged even as the demand for labor (i.e., the number of individuals required to handle those tasks) changes. On the other hand, even if the demand for a given role remains constant, a single technology-induced change in worker productivity can carry multiple, competing implications for the role's cognitive and social skill demand.

This is because GenAI can affect the importance of a focal skill in a role directly, by adding (or eliminating) tasks that draw on that skill, and indirectly, by adding (or eliminating) other tasks that draw on different skills and thereby shifting the relative weight of those that remain. The importance of social skills within a role may thus decline (rather than increase, as usually argued) if GenAI eliminates the interaction-related tasks assigned to a worker. It may also decline through the indirect channel alone, if GenAI adds tasks drawing on other skills, reducing social skills' share of the role even when interaction-related tasks are never directly affected. Finally, the relative demand for social skills may also decrease if GenAI enables workers with lower social-skill proficiency to perform interaction-related tasks effectively, reducing the level of social skill required even when the tasks themselves remain unchanged. Without careful consideration of these alternatives, the conjecture that GenAI amplifies the importance of social skills remains, at best, an incompletely specified prior.

We therefore build on previous conceptual models characterizing the effects of technology change on tasks (e.g., Acemoglu et al., 2025; Acemoglu & Restrepo, 2019) to develop a skill-centric framework (Lazear, 2009; Raveendran et al., 2022) that can help interpret the mechanisms underlying observed changes in skill requirements for a role following the adoption of GenAI. This framework identifies the conditions under which the intuition about the increasing importance of social skills and the decreasing relevance of cognitive skills following GenAI adoption in a given role is sound and when it is not.

Empirically, we investigate changes in the stated demand for cognitive and social skills, relative to other skills, in roles that explicitly begin to require GenAI use (e.g., "prompt engineering," or familiarity with tools such as ChatGPT, Google Gemini, or Anthropic Claude, to name a few; henceforth, GenAI roles). Our approach rests on the assumption that changes in the relative emphasis on skills required for a role, in a difference-in-differences (henceforth, DiD) framework, provide evidence about how GenAI adoption reshapes the human capabilities required to perform in each role. We do not investigate whether GenAI adoption alters the distribution of roles themselves in firms but instead study changes within persistent roles after GenAI-related keywords have been incorporated within them.

We use job postings from Lightcast to study within-role changes in the relative demand for cognitive and social skills following the launch of ChatGPT, a popular GenAI tool released on November 30, 2022, and subsequently adopted widely across firms. Our dataset includes 6.961 million postings from 595 U.S. public firms covering 214 six-digit Standard Occupational Classification (SOC) occupations. Lightcast classifies these postings into 41,852 fine-grained roles (e.g., "software engineers," "DevOps engineers," and "solutions architects" are Lightcast roles that map to SOC code "15-1252.00 software developers"), crucially allowing us to track skill requirements within roles over time, rather than across occupations.

GenAI roles are identified using a keyword approach, which we validate with independent survey- and model-based indicators (Bick et al., 2024; Eloundou et al., 2023). Cognitive and social skill intensity, i.e., the demand for such skills, relative to the total number of skills in each posting, are measured based on the count of skills appearing in the respective dictionaries created by Deming and Kahn (2018), which we also benchmark against publicly available skill intensity ratings from the occupational information network (O*NET).

Using firm-role-month observations, we estimate DiD models comparing GenAI and non-GenAI roles before and after ChatGPT's launch, with firm-role and month fixed effects and standard errors clustered at both the firm and role levels. This allows us to document changes in stated skill portfolios within GenAI roles relative to non-GenAI roles following ChatGPT's release, isolating shifts in skill composition from role creation and elimination.

We find that GenAI roles exhibit a 0.111 percentage point decline in social-skill intensity (3.392% of the sample mean) relative to comparable non-GenAI roles after ChatGPT's launch, while cognitive-skill intensity shows no detectable change in the observation period. Further, the decline in social skill intensity is not driven by an absolute decline in social skills (the count of social skills remains stable within GenAI roles). Rather, it appears to be driven by an absolute increase in other skills, specifically functional administrative skills, i.e., those related to finance, human resources, and legal functions. These patterns are consistent with what our framework describes as skill reweighting.

Additional analyses show that this increase in functional administrative skills is concentrated in non-managerial occupations outside the corresponding functions, suggesting that skill reweighting is driven primarily by cross-functional skill extension. This is consistent with two alternative interpretations: GenAI adoption in a role may intensify coordination with administrative functions, or it may enable self-service of tasks previously handled by those administrative functions. Because the increase is strongest in roles that are less coordination-oriented (i.e., non-managers), we conclude that the self-service explanation appears more plausible.

This paper makes three contributions to research on the relationship between technology adoption and the nature of work (e.g., Acemoglu & Autor, 2011; Adner et al., 2019; Bresnahan, 2011). First, while prior work has examined how GenAI affects roles and tasks (e.g., Dell'Acqua et al., 2023; Handa et al., 2025), we develop a skill-centric framework that clarifies the conditions under which technology adoption increases or decreases the relative importance of a given skill within a role and shows why the common expectation of social skill amplification following GenAI adoption is theoretically contingent rather than inevitable. Second, we provide early empirical evidence of how GenAI adoption reshapes skill requirements within roles, showing that social skill intensity in GenAI roles declines because roles broaden to include additional cross-functional administrative skills. Third, by showing that this cross-functional skill addition is concentrated in non-managerial roles and occurs without a corresponding increase in social skills, our results suggest that GenAI's early labor-market impact may be less about intensifying coordination and more about enabling workers to perform tasks previously handled by those functions.

## 2. PRIOR LITERATURE AND THEORETICAL FRAMEWORK

### 2.1. GenAI adoption and the nature of work

Since the release of ChatGPT in November 2022, the development and adoption of GenAI technologies have progressed rapidly. GenAI refers to a class of deep-learning models capable of producing new content (e.g., text, voice, music, images, and video) based on patterns learned from training data (IBM Research, 2023). In this paper, we focus on a specific subset of GenAI tools: large language models (LLMs), such as ChatGPT released by OpenAI, trained on extensive text corpora drawn from sources including websites, social media, articles, and books. These systems enable users to perform day-to-day tasks through interactive, conversational prompting. A key distinction between LLMs and earlier automation technologies is that their capabilities extend beyond tasks with explicit step sequences that can be codified and programmed, allowing for the automation of non-routine tasks previously viewed as uniquely performable by humans (Acemoglu & Restrepo, 2019; Argote et al., 2021).

Several studies have investigated the impact of GenAI on the nature of work. In previous research, a *role* is conceptualized as a collection of *tasks*, with individuals occupying the role bringing to bear a set of *skills* relevant to those tasks (Acemoglu & Autor, 2011; Lazear, 2009; Raveendran et al., 2022). The number of incumbents in a role equals the number of *jobs* associated with it. *Occupations* are families of related roles.

Prior research on GenAI and work falls into three broad categories. The first examines the expected exposure of work tasks and occupations to GenAI (Bick et al., 2024; Eisfeldt et al., 2023; Eloundou et al., 2023; Felten et al., 2023), finding that a substantial share of job roles have significant task exposure to GenAI (Eloundou et al., 2023) and that high firm-level GenAI exposure correlates positively with stock market value following the launch of ChatGPT (Eisfeldt et al., 2023). The second leverages field experiments to assess GenAI adoption in roles such as customer service agents (Brynjolfsson et al., 2023), consultants (Dell'Acqua et al., 2023), entrepreneurs (Otis et al., 2023), and writers (Doshi & Hauser, 2024). Their findings consistently indicate productivity gains in such professions post-GenAI adoption. The third set (e.g., Handa et al., 2025; Hoffmann et al., 2024) analyzes tasks, and uses natural experiments and GenAI-human interactions to examine how GenAI adoption simultaneously enhances and displaces specific tasks within given individual workflows.

This paper extends prior research by shifting the focus from changes in tasks and roles (i.e., bundles of tasks) to changes in the skills demanded following GenAI adoption. Skills are defined as the developed mental and physical capacities required to perform tasks, and tasks are defined as units of work activity that produce a particular outcome (O*NET Online, 2025). Skills constitute the most granular and foundational building block of work. While some skills are task specific (e.g., programming in Python or performing capital budgeting), others are not tied to any single task but instead support the performance of a wide range of them (Hilton & Pellegrino, 2012). Conversely, a task may draw on multiple skills, such that tasks and skills have a many-to-many, rather than one-to-one, mapping. Yet, existing research often adopts a level of analysis that is too coarse to capture the impact of GenAI adoption on the demand for these cross-cutting skills. In particular, a task and role level focus risks misestimating changes in skill demand when tasks remain the same, but skills requirements change, or when tasks change but rely on the same skills.

We focus on cognitive and social skills, i.e., domain-agnostic, foundational capabilities that support performance across a wide range of tasks, for two reasons. First, these skills are central to effective performance and to adaptation when tasks evolve (e.g., Dimitriadis & Koning, 2022; Lewis & Smith, 1993). Second, because the mapping between tasks and social and cognitive skills is not one to one, evidence on task-level change drawn from prior literature is insufficient to infer how the demand for these skills shifts following GenAI adoption. Our paper attempts to address this gap.

**2.2. The importance of cognitive and social skills**

Cognitive skills involve workers' capacity for problem solving, critical thinking, analysis, creativity, and decision making, among others (Deming & Kahn, 2018). They allow for higher-order thinking, which is the capacity to integrate new and stored information, recombine it, and apply it purposefully to novel situations (Lewis & Smith, 1993). While individuals vary in their innate cognitive endowments, these skills can be significantly improved through education, training, and particularly through working on "stretch problems" just beyond one's existing capabilities (Lewis & Smith, 1993). In doing so, individuals not only improve current performance but also expand their capacity for future learning.

The dual benefit of enhancing both current performance and future learning capacity makes cognitive skills especially valuable as automation handles routine tasks and workers focus on non-routine analytical work requiring judgment and creativity (Deming, 2022). Meta-analyses consistently show cognitive ability as the strongest predictor of job performance (Schmidt & Hunter, 1998), with recent evidence suggesting decision-making skills now predict labor income independently of education and IQ (Caplin et al., 2025).

Social skills, in turn, enable workers to interact with others and encompass communication, collaboration, negotiation, presentation, and teamwork (Deming, 2017). These skills enhance productivity in collaborative tasks by helping individuals anticipate coworkers' needs, communicate more effectively, manage conflict, and coordinate their work. Critically, they also accelerate learning, as individuals acquire much of their knowledge through observing, questioning, and collaborating with others. Capacities for clear communication, perspective taking, and empathic listening become, thus, central to skill development. Research demonstrates that teams with strong social skills share information more efficiently, surface errors sooner, and recombine ideas more effectively (Dimitriadis & Koning, 2022; Hoffman & Tadelis, 2021; Klein et al., 2006; Mayer et al., 2008).

In sum, cognitive and social skills are domain agnostic, foundational, and especially relevant to non-routine tasks that require judgment and/or interpersonal coordination (Anderson & Krathwohl, 2001; Lewis & Smith, 1993). These skills enable workers to adapt to changing task requirements and, accordingly, provide a basis for continued employability, especially when technological disruption renders existing tasks obsolete and creates new ones.

In this sense, they can be viewed as "meta skills," the individual-level analog of firm-level dynamic capabilities (Teece et al., 1997; Winter, 2003). Further, since firm-specific human capital (Becker, 1962) rests on learning how to work within the specific task environments of a firm, cognitive and social skills are also important inputs for firms seeking to build competitive advantage on this basis (Wang & Barney, 2006). Reflecting this importance, demand for cognitive and social skills has grown steadily over recent decades (Deming, 2017) and annual surveys consistently find that employers cite them as among the most important and hard-to-fill competencies for workers (Cunningham & Villaseñor, 2016; NACE, 2024).

Since cognitive and social skills and GenAI are both central to non-routine task performance, understanding how GenAI adoption and usage reshape demand for these skills is critical. To do so, in this paper, we introduce a skill-centric framework for assessing technology-driven change, building on prior work on tasks (e.g., Acemoglu et al., 2025; Acemoglu & Restrepo, 2019). We then apply this framework to empirically examine how GenAI adoption relates to changes in the relative demand for cognitive and social skills stated in job postings.

### 2.3. Proposed theoretical framework: A skill-centric view of technological change

A task-centric perspective highlights three distinct ways in which technologies can reshape work tasks. First, they may increase the productivity of human effort at a task, allowing workers to generate more output with the same effort, acting as a multiplier on their skills (complementarity). Second, they may fully displace tasks by automating them (displacement). Third, they may create new tasks, either because the technology introduces new productive possibilities that require human oversight or because it makes previously infeasible or low-value activities worthwhile (task creation). These effects are well established in existing theories of technological change and labor demand (Acemoglu et al., 2025).

In the skill-centric framework we propose, we take one additional step: Rather than focusing on how technology changes tasks, we examine how these task-level effects translate into changes in the demand for skills within roles. There are two reasons why focusing on the skill level, rather than the task level, is important. First, skills are the capabilities, grounded in knowledge and experience, that create value for both workers and firms. For employees, skills are what they are compensated for and what sustain their employability (Lazear, 2009). For firms, hiring is fundamentally an exercise in acquiring skills, which act as an effort multiplier and, therefore, underpin competitive advantage. Even when two firms perform the same set of tasks, differences in their workforces' skills can result in systematically different outputs. Second, the relationship between skills and tasks is rarely one to one: Tasks typically draw on multiple skills, and the same skill can be deployed across multiple tasks (Raveendran et al., 2022). As a result, an exclusive focus on tasks may misestimate a technology's impact on work: tasks can remain stable while the underlying skill requirements shift, and the very same task change can move the relative demand for a given skill in opposite directions, depending on how intensively that task drew on a focal skill relative to the others it bundles together.

In building the skill-centric framework, we follow standard assumptions in the literature on the impact of technology on work (e.g., Acemoglu et al., 2025; Acemoglu & Restrepo, 2019). We assume work in firms to be organized around roles, i.e., bundles of tasks that an individual is expected to perform. Tasks require workers to apply skills and effort (Lazear, 2009; Raveendran et al., 2022). Importantly, however, we assume that a single task may draw on multiple skills (e.g., coordinating a technical implementation requires both technical and social skills) and a single skill may be applied across multiple tasks within a role (e.g., analytical reasoning may be required for both data analysis and strategic planning). This many-to-many mapping between tasks and skills is what distinguishes a skill-centric from a purely task-centric view.

We start by defining *relative skill demand* within a role, i.e., the importance of a focal skill $k$ relative to the role's full set of required skills. Empirically, this corresponds to the intensity $R$ (i.e., the share of total required skills) of skill $k$ in role $r$:

$$R_{kr} = \frac{s_{kr}}{\sum_j s_{jr}} \qquad (1)$$

where $s_{kr}$ is the requirement for skill $k$ in role $r$, and $\sum_j s_{jr}$ aggregates the requirements for all skills $j$ listed for the role $r$. The task-centric mechanisms of complementarity, displacement, and task creation (Acemoglu et al., 2025) discussed above translate into changes in relative skill demand $R_{kr}$ through (a) tasks in role $r$ that load on skill $k$ changing importance, thus shifting the numerator $s_{kr}$, and/or (b) requirements for skills in role $r$ *other than* $k$ rising or falling as associated tasks change importance, thus shifting the denominator $\sum_j s_{jr}$.

The relative skill demand can decrease within a role under three conditions. First, the numerator in equation (1) may decline through task deskilling, i.e., when the technology reduces the skill proficiency required to perform tasks in the role, allowing lower-skilled employees to perform tasks that previously required higher expertise (e.g., GenAI automating complex analysis). This can reduce the stated requirement for skill $k$, leading to a decline in $s_{kr}$, despite the technology not altering the set of tasks associated with the role.

Second, the requirement for skill $k$ in role $r$ $s_{kr}$, i.e., the numerator in equation (1), may decline through skill displacement. Specifically, when the technology displaces tasks that draw on skill $k$, the role's reliance on skill $k$ decreases, leading to a reduction in $s_{kr}$ . Note also that because a single task may draw on multiple skills, displacing one task can simultaneously reduce the demand for several skills.

Third, the total requirements across all skills listed for role *r,* $\sum_j s_{jr}$, i.e., the denominator in (1), can increase through skill reweighting. This occurs when a role's requirements for other skills rise, either because new tasks are added that draw on those skills, or because existing tasks shift toward greater reliance on them. In this case, $R_{kr}$ falls mechanically even if $s_{kr}$ remains unchanged. Unlike the first two mechanisms, skill reweighting does not require the technology to have a direct effect on skill $k$ itself.

Conversely, the relative demand for skill $k$ can increase within a role under three mirror-image conditions. First, upskilling occurs when the technology complements skill $k$ by raising the proficiency required to perform tasks in the role, so that tasks previously within reach of lower-skilled workers now demand higher expertise, thus raising $s_{kr}$. Second, skill expansion occurs when the technology creates new tasks that draw on skill $k$, directly increasing $s_{kr}$. Finally, skill reweighting operates through the denominator, $\sum_j s_{jr}$: When requirements for other skills fall, because tasks drawing on those other skills are either deskilled or displaced, $R_{kr}$ rises mechanically even if $s_{kr}$ does not change.

Table 1 illustrates the conceptual framework described above through a selection of representative cases. By explicitly linking technology's effects on tasks to changes in the relative demand for skills, our framework clarifies why the task-centric perspective developed in prior work cannot, on its own, account for the effects of technology adoption on skill demand. In fact, the canonical task-level effects of technology (complementarity, displacement, and creation) do not map one to one onto skill-level effects (upskilling/deskilling, skill displacement/expansion, and skill reweighting). Indeed, the same task-level change can imply opposite shifts in relative demand for the skills that the task draws on, as illustrated by the examples in Table 1. A technology that leaves a task intact can still generate either deskilling or upskilling, depending on whether productivity gains are captured via lower proficiency requirements or via expanded output. Likewise, task creation can manifest as either skill expansion or skill reweighting, and task elimination can also lead to either skill displacement or skill reweighting.

--- INSERT TABLE 1 ABOUT HERE ---

The proposed framework also shows why the common view that GenAI automates cognitively intensive work and thereby makes social skills more relevant cannot be unconditionally correct. This view implicitly assumes that GenAI (1) does not deskill or displace social skill-intensive tasks (so the social-skill numerator in (1) does not fall) and (2) does not expand requirements for other non-social skills (so the denominator does not rise), either of which would offset or reverse the effects of any downward shift in cognitive skills.

While the relative-demand framework helps parsimoniously describe the channels through which skill demand may shift following the adoption of a work-altering technology, it cannot unambiguously determine which channel – upskilling/deskilling, skill displacement/expansion, and skill reweighting – drives the change in a skill's relative demand. For example, if we observe a reduction in the relative demand for writing skills following GenAI adoption, this is consistent with skill displacement, if GenAI removes tasks that relied on writing, or skill reweighting, if GenAI shifts employees toward additional tasks that draw on other skills – reducing the relative demand for writing. These mechanisms may also operate simultaneously, in which case the net effect does not reveal which one dominates. Separating them requires examining changes in absolute skill demand – i.e., the numerator $s_{kr}$ of the relative-demand measure – alongside relative demand, which we also do in this paper.

Also, the framework does not yield a priori predictions about GenAI's effects on cognitive and social skills. The relative magnitude of the underlying mechanisms is unknown, as are the resulting net effects on skill demand. The framework is nonetheless useful for interpreting empirical patterns by ruling in or ruling out plausible mechanisms. Accordingly, we take a data-driven approach to the problem.[1] We first document changes in cognitive and social skill demand in roles that adopt GenAI and then use the framework to infer mechanisms consistent with the observed effects.

[1] See Graebner et al. (2023), Marchetti and Puranam (2026), Nary (2024), and Sen and Puranam (2022) for similar empirical investigations.

# 3. DATA AND METHODS

## 3.1. Data source and sample

To study the impact of GenAI adoption on within-role changes in the relative demand for social and cognitive skills, we leverage data from Lightcast (previously, Burning Glass Technologies). Lightcast collects, aggregates, and deduplicates jobs posted by firms online from ~220,000 sources. For each opening, Lightcast collects the job posting's full text, the posting date, and the hiring employer's name. Lightcast also leverages the job postings' raw text to generate variables such as the aggregate role type, O*NET occupation mapping, and the skills required for each role, among others, which we use in this paper. The main advantage of these data is that they offer time-varying and granular insights into the demand for skills as firms adapt to fast-evolving technologies such as GenAI.

The Lightcast data map job postings onto roles and standardized occupations, providing a classification that is particularly well suited for studying changes in the stated relative demand for cognitive and social skills in GenAI roles, i.e., roles in which workers are expected to use GenAI to perform their tasks, as detailed in Section 3.3 below. Job postings represent the lowest level of granularity in this categorization, followed by Lightcast roles, which are, in turn, more fine grained than O*NET SOC codes. For instance, data scientist, data analyst, and business intelligence engineers are different Lightcast roles that are mapped onto the same data scientist O*NET SOC. Appendix A provides more examples and further details on the mapping between Lightcast roles and O*NET SOCs. In the rest of the paper, we refer to the Lightcast classification as "roles" (i.e., an aggregation of jobs posted by the firm) and to the O*NET six-digit SOC classification as "occupations" (i.e., an aggregation of roles), consistent with the theoretical framework in Section 2.1.

To build our sample, we start by identifying job postings for GenAI roles advertised by U.S. public firms[2] between January 2022 and December 2024. The selected time window allows us to compare the impact of GenAI on roles before and after its adoption, spanning more than 24 months after the launch of ChatGPT in November 2022. We identify GenAI roles using a dictionary search of the job posting text, as detailed in Section 3.3, consistent with previous research on firm technology adoption (e.g., Alekseeva et al., 2024; Goldfarb et al., 2023; Gulati et al., 2023). This approach yields a sample of 29,587 job postings for GenAI roles advertised by 595 U.S. public firms since November 2022 and corresponds to 3,569 roles and 214 occupations. We use this measure to identify firms that have adopted GenAI. Figure 1 shows the monthly distribution of advertised GenAI roles in the sample.

--- INSERT FIGURE 1 ABOUT HERE ---

Subsequently, we merge our sample of 29,587 GenAI job postings with 6,931,885 other job postings by the same firms that do not include the GenAI keywords but belong to the same 214 occupations previously identified in the period between January 2022 and December 2024. These non-GenAI job postings correspond to 38,283 Lightcast roles in the same 214 occupations the GenAI roles map onto. Our final sample comprises 6,961,472 job postings across 41,852 roles (of which 3,569 adopt GenAI and 38,283 do not), 214 occupations, and 595 firms from January 2022 (10 months before the ChatGPT launch month, and after the COVID pandemic-related business disruptions had subsided[3]) to December 2024 (26 months including the ChatGPT launch month). For the main analyses, we aggregate the posting-level data to the firm-role-month level by averaging the measures within each firm-role-month cell, yielding 1,709,731 firm-role-month observations.

---

[2] We restrict our sample to public firms because of data reliability and the possibility of accessing additional information about the same firms for validation and triangulation.

[3] Between May 2020 and December 2021, the share of employees teleworking specifically because of the COVID-19 pandemic fell from 35.4% to 11.1%, while the share of employees unable to work because their employer had closed or lost business fell from 19.2% to 1.2%; thereafter, the decline was more gradual: 5.2% and 0.5% by September 2022 (Source: https://www.bls.gov/cps/covid-may2020-sept2022-data-tables.htm).

### 3.2. Dependent variables

To assess changes in the *relative demand* for cognitive and social skills following the adoption of GenAI, we calculate *cognitive* and *social skill intensity* using Deming and Kahn's (2018) skill categorization in conjunction with the Lightcast taxonomy. Deming and Kahn's (2018) categorization comprises keyword dictionaries compiled by the authors to capture cognitive and social skills.[4] The Lightcast skill taxonomy (e.g., Burning Glass Technologies, 2019; Goldfarb et al., 2023) consists of approximately 17,000 algorithmically identified standardized keywords corresponding to individual skills and maps each job posting to a subset of these keywords that reflects its skill requirements.

To measure the dependent variables, we start by manually identifying Lightcast skills that match the Deming and Kahn's (2018) keyword dictionaries for cognitive and social skills. These mappings are available in Appendix B. Next, we calculate a job posting's *cognitive* and *social skill intensity* as the ratio of the number of cognitive or social skills to the total number of skills listed for the job posting. A higher skill intensity indicates that the job posting places a greater relative demand on those skills than on other skills. For additional analyses, as reported in Section 4.4, we also calculate *cognitive* and *social skill count* to measure their absolute demand by counting the number of skills corresponding to the cognitive and social skill categories for the focal job posting and average these measures at the firm-role-month level.

Our measures of *cognitive* and *social skill intensity* assume that job postings reflect demand for skills in a role within organizations. While measurement error is possible in our operationalization, we conduct validation tests to support this assumption. Specifically, we compare the *cognitive* and *social skill intensity* measures to skill-level ratings from O*NET, which are based on surveys of workers employed in occupations covered within the O*NET categorization and describe the corresponding work attributes and job characteristics.[5] We find a positive correlation between the proposed Lightcast-posting-based and the O*NET-worker-survey-based skill intensity measures (0.537 for *cognitive skill intensity* and 0.502 for *social skill intensity*), supporting the assumption that skill intensity measures based on job postings reflect role characteristics. Appendix B provides additional details on the construction and validation of the dependent variables.

---

[4] Deming and Kahn (2018) identify the following keywords as indicative of cognitive skills: analytical, critical thinking, math, problem solving, research, and statistics, and the following keywords as indicative of social skills: collaboration, communication, negotiation, presentation, and teamwork.

### 3.3. Independent variable

To measure GenAI adoption, we start by identifying job postings that reference GenAI technologies. Because GenAI entered commercial use only in late 2022, no standardized taxonomy for GenAI job content yet exists. Following prior work that measures technology adoption using a dictionary-based text search (e.g., Alekseeva et al., 2024; Goldfarb et al., 2023; Gulati et al., 2023), we construct a keyword dictionary through iterative review of job postings in our sample.

Our dictionary comprises three types of terms: (1) generic technology references ("generative AI," "conversational AI," "LLM," "large language model"), (2) named products and platforms ("ChatGPT," "GPT-3," "GPT-4," "Microsoft Copilot," "GitHub Copilot," "Google Bard," "Google Gemini," "Mistral AI," "Meta LLaMa," "Anthropic Claude," "Perplexity AI," "Grok," "Deepseek"), and (3) role descriptors specific to GenAI workflows ("prompt engineer"). We deliberately restrict the dictionary to terms that unambiguously reference GenAI, excluding broader terms (e.g., "artificial intelligence," "machine learning," "NLP") that predate and are not specific to the generative paradigm. This approach, combined with the fact that some roles may apply GenAI without explicitly mentioning it in the corresponding job ads, likely yields a conservative estimate of GenAI adoption across the work roles in our sample. As a validation check, we manually reviewed a random sample of postings to assess whether the keywords appeared in contexts related to GenAI use; Appendix C presents snapshots of two illustrative examples.

[5] We use survey responses collected by O*NET under the O*NET Data Collection Program (approximately every 5 to 7 years). For each occupation, O*NET surveys workers and occupational experts about work attributes and job characteristics. Measures of social and cognitive skills constructed from O*NET therefore provide a useful benchmark to validate our job-posting-based measures. Appendix B lists the subset of survey questions used to construct the O*NET-based social and cognitive skill measures.

Using this list of keywords, we define GenAI adoption at the role level, since we are interested in measuring how such roles evolve over time in their relative skill demand compared to others within the same firm. The variable *GenAI role* takes the value of 1 if the text of at least one job posted within the focal role after November 2022 includes one of the keywords above (3,569 roles in our sample) and 0 otherwise. We also define *post ChatGPT* as a dichotomous variable set to 1 for each month from November 2022 (i.e., the launch month of ChatGPT) onwards and 0 otherwise to capture the ChatGPT release timeline. This operationalization implies that we treat GenAI adoption as a non-staggered exogenous treatment. We discuss this identifying assumption further in Section 3.4.

To validate our measure, we compare the measure of GenAI adoption constructed from job postings to two independent benchmarks that provide occupation-level adoption measures: (1) worker survey evidence on GenAI use (Bick et al., 2024) and (2) predictive exposure estimates based on task content (Eloundou et al., 2023). Our measure correlates strongly with both benchmarks (0.720 with the worker-survey adoption measure and 0.724 with the exposure-based measure).[6] Appendix C provides details on these validation tests.

[6] For each two-digit SOC occupation group, we compute the share of job postings containing at least one GenAI-related keyword. We use only 2024 postings for this validation, because the occupation-level adoption rates in Bick et al. (2024) were collected in 2024. We then compare our 2024 posting-based adoption measure with two external benchmarks reported in Figure 8 of Bick et al. (2024): their occupation-level adoption rates and Eloundou et al.'s (2023) predicted-exposure measure.

### 3.4. Estimation strategy

For the main analysis (results reported in Sections 4.3 and 4.4), we employ a DiD estimation strategy paired with an ordinary least squares (OLS) specification model, as follows:

$$y_{ijt} = \alpha_{ij} + \lambda_t + \beta * GenAI\ Role_i * PostChatGPT_t + \epsilon_{ijt} \quad (2)$$

where $y$ is either *cognitive* or *social skill intensity* or *absolute skill count* for role $i$, within firm $j$ and month $t$. The estimation model in Equation (2) also includes firm-role fixed effects ($\alpha_{ij}$) and month fixed effects ($\lambda_t$). We are interested in the sign and magnitude of the coefficient β that estimates the changes in the intensity of cognitive and social skill demand for GenAI roles (on average across job postings) after ChatGPT launch as compared to other roles. Since roles differ in the number of jobs advertised, in our estimation, we place more importance on roles that advertise more jobs by weighting each firm-role-month level observation by the count of jobs posted within that observation unit. Since skill demand may be correlated across roles within the same firm and across firms advertising for the same role, we cluster standard errors at both the firm and role levels.

The proposed empirical design assumes that in November 2022, all roles suitable for GenAI adoption were revealed through the exogenous release of ChatGPT. Our approach identifies these *GenAI roles* as those with at least one job posting by any firm in our observation window containing GenAI-related keywords.[7]

## 4. ANALYSES AND RESULTS

Our analysis proceeds in three steps. We first describe where GenAI roles have emerged across occupations and industries (Section 4.1). We then examine, in cross section, whether GenAI roles differ systematically from other roles within the same occupation and firm to establish whether selection into GenAI adoption correlates with pre-existing skill profiles (Section 4.2). Finally, we turn to our main DiD analysis, in which we track changes in skill demand within the same role and firm over time, comparing GenAI roles to non-GenAI roles before and after ChatGPT's launch (Sections 4.3-4.4). In other words, the cross-sectional analysis compares across roles within an occupation to establish baseline differences, and the DiD analysis exploits within-role variation over time to estimate the effect of GenAI adoption on skill demand.

[7] An alternative would have been to estimate staggered adoption models that take the first occurrence of GenAI-related keywords in a role within a firm as the time of treatment. However, given our shallow window of observation, this would significantly diminish the instances of adoption, and we believe it is reasonable to ignore staggering (across firms and roles) in such a relatively short period and assume that roles that are suitable for adopting GenAI are all in the treatment group from November 2022, and different from roles that never adopt during our observation period (i.e., the control group).

### 4.1. Descriptive analyses: Where have GenAI roles emerged?

Before analyzing the effect of GenAI adoption on the relative demand for cognitive and social skills, we begin by describing the occupations and industries in which GenAI roles have emerged. Figure 2 shows the top five occupations and industries with the highest percentage of GenAI roles. Computer and mathematical occupations account for the largest share of GenAI roles (54.454%). This aligns with expectations, as software developers are likely among the first types of roles required by their firms to use the technology, given its nature. Other research supports this trend, showing that these roles are the most likely to adopt GenAI, as indicated by both predictive exposure calculations (Eloundou et al., 2023) and worker surveys (Bick et al., 2024). In terms of industries, retail trade (29.367%)[8] and information technology (19.169%) industries have the highest rate of adoption.

--- INSERT FIGURE 2 ABOUT HERE ---

### 4.2. Cross-sectional analyses: Cognitive and social skill intensity of GenAI roles before ChatGPT

Before estimating the effect of GenAI adoption on skill demand, we characterize how GenAI and non-GenAI roles differ in their cognitive and social skill intensity prior to ChatGPT's launch. Understanding these baseline differences will help inform the interpretation of our DiD estimates. Specifically, we compare the *cognitive* and *social skill intensity* across GenAI and non-GenAI roles within the same firm and occupation, restricting the sample to the pre-launch period (January-October 2022). We estimate OLS models with firm-occupation and month fixed effects and cluster standard errors at the firm and occupation levels.[9]

[8] Among retail trade firms, Amazon.com Inc is the firm with the highest percentage of GenAI roles advertised.

Figure 3 plots the results of this analysis. We find that *cognitive skill intensity* (e.g., problem solving, critical thinking, etc.) is 0.201 percentage points higher ($p = 0.022$) in GenAI roles than in other roles prior to ChatGPT launch. Relative to the sample mean, reported in Table 2, the value is 6.852% higher. For social skills, the difference between GenAI roles and other roles is statistically insignificant ($\beta = 0.046$, $p = 0.586$). These results suggest that roles adopting GenAI were already more cognitively intensive than comparable roles (within the same firm and occupation) before ChatGPT's launch, while their social skill intensity was indistinguishable. This asymmetry provides a useful baseline for interpreting the DiD estimates that follow: Any post-launch changes in cognitive skill demand occur against an already elevated baseline, whereas changes in social skill demand would represent a departure from parity.

--- INSERT FIGURE 3 AND TABLE 2 ABOUT HERE ---

### 4.3. DiD analyses: Changes in cognitive and social skill demand in GenAI roles following ChatGPT's launch

To address the main research question – how the cognitive and social skill intensity of GenAI roles evolve over time – we compare changes in GenAI roles post-ChatGPT launch (November 2022) to the pre-launch period, using other roles posted by the firm as the counterfactual group. Table 3 (models 1-2) and Figure 4 (panel 1 "Skill Intensity") present the results of this analysis. We find a decrease in the intensity of social skills listed in GenAI roles compared to the pre-ChatGPT launch period, adjusting for changes in the counterfactual group and controlling for firm-role and month fixed effects. As shown by the red line in the first panel of Figure 4, representing changes in *social skill intensity*, the value decreases by 0.111 percentage points ($p$ = 0.006) relative to the pre-launch period. Relative to the sample mean, the value is 3.392% lower. For *cognitive skill intensity* (blue line), on the other hand, we do not observe any statistically significant change ($\beta$ = -0.050, $p$ = 0.681). Time-series estimates, calculated for each month in the sample and plotted in Figure 5(a), further illustrate these findings. We observe that, post-ChatGPT launch, monthly social skill intensity estimates decrease over time, especially towards the end of our short observation window (after 18 months). In contrast, pre-launch estimates are statistically indistinguishable from zero, indicating parallel trends pre-ChatGPT launch.

[9] This differs from our main specification, where we use firm-role fixed effects and cluster standard errors at the firm and role levels, because, for this analysis, we rely on cross-sectional variation across roles.

The decline in *social skill intensity* within GenAI roles is consistent with (1) deskilling, when GenAI reduces the social skill proficiency required to perform tasks in a role, (2) skill displacement, when the technology displaces tasks requiring social skills from the role, and (3) skill reweighting, an increase in the importance of other skills in the role. To tease out these alternative mechanisms driving the decline in *social skill intensity*, we further analyze the impact of GenAI adoption on the absolute skill count, rather than on skill intensity (i.e., the relative skill count).

We re-estimate our models using the inverse hyperbolic sine transform of absolute skill counts for cognitive, social, and other skills rather than relative ratios. The skill category "other skills" includes all skills tagged in job postings that do not correspond to cognitive and social skills. Table 3 (models 3-5) and Figure 4 (panel 2 "Skill Count") present results for this analysis. We find no change in social skill counts ($\beta$ = -0.008, $p$ = 0.391) and a simultaneous significant increase in other skill counts ($\beta$ = 0.024, $p$ = 0.031), both contributing to a statistically significant decline in social skill intensity. Time-series estimates plotted in Figure 5(b) further illustrate these findings.

Taken together, the decline in social-skill intensity, the absence of change in the total number of required social skills, and the increase in the number of other skills suggest skill reweighting within roles. This pattern implies that GenAI adoption broadens the skill requirements of a role beyond social and cognitive skills rather than displacing existing social or cognitive skill demands.

--- INSERT FIGURES 4 AND 5 AND TABLE 3 ABOUT HERE ---

## 5. ROBUSTNESS ANALYSES

We conduct several robustness checks to assess the validity of our results. First, as the DiD estimates could reflect changes in either the treated group (GenAI roles) or the counterfactual group (other roles), we also examine within-group trends over time. Specifically, we plot changes in outcomes separately for GenAI roles and other roles, controlling for firm-role fixed effects. We focus on *social skill intensity* and *other skill count*, the outcomes for which we observe statistically significant effects in the main analysis. The corresponding figures are reported in Appendix D. Two patterns emerge: (1) Social skill intensity rises across both GenAI and non-GenAI roles, but the increase is markedly smaller for GenAI roles, producing the divergence captured by our DiD estimates, and (2) consistent with the main results, this divergence is driven by a larger increase in the number of other skills in GenAI roles relative to other roles. The effects become pronounced towards the end of our (relatively short) observation window. Together, these patterns corroborate our main findings and are consistent with a skill reweighting interpretation.

Second, because we find a decline in social skill intensity, we examine whether it could be mechanically driven by the tagging of GenAI-related skills in job postings after November 2022, since this risks inflating the total number of skills tagged to GenAI role postings and the denominator of the social skill intensity measure. As a robustness test, we thus exclude from the skill taxonomy skills corresponding to the keywords used to identify GenAI roles and are among the most frequently tagged skills after November 2022, while being absent before then.[10] Their removal thus produces the most significant reduction in any mechanically induced increase in the skill intensity denominator. We then recompute skill-intensity and skill-count measures. Our main results remain robust to this exclusion, as reported in Appendix Table E1.

Third, because our argument centers on GenAI use rather than GenAI production or sales, we ensure that the results are not driven by firms directly involved in developing LLMs or providing related consulting services. Accordingly, following Acemoglu et al. (2022), we exclude firms in the information sector (NAICS 51) and the professional and business services sector (NAICS 54). As reported in Appendix Table E2, our main results on *social skill intensity* remain robust to the exclusion of these sectors ($\beta = -0.106$, $p = 0.005$). We also find a similarly large and positive increase in the absolute count of other skills, although the estimate is accompanied by a larger p-value ($\beta = 0.020$, $p = 0.103$). Relatedly, we also verify that our findings are not concentrated among technical roles, where shifts in skill demand might reflect the production of GenAI technologies rather than their adoption. We therefore exclude technical roles, i.e., those mapped to “15-computer and mathematical” occupations in the O*NET SOC classification, and report results’ robustness in Appendix Table E3.

## 6. SUPPLEMENTARY ANALYSES

The results reported in Section 4 point to a skill reweighting effect of GenAI adoption, as we observe a decline in social skill intensity, no change in the total number of required social skills, and an increase in the number of other skills (i.e., all skills except social and cognitive ones). To shed further light on these results, we examine two additional related questions: (1)

[10] The list of Lightcast skills removed in this analysis are: “generative artificial intelligence,” “ChatGPT,” “prompt engineering,” “retrieval augmented generation,” “Azure OpenAI,” “Microsoft Copilot,” “AI agents,” and “GitHub Copilot.”

which specific skills contribute to the observed reweighting effect and (2) which roles drive the main results?

### 6.1. Skills contributing to the observed skill reweighting effect

To identify which specific skills contribute to the skill reweighting effect that drives the observed decline in social skills intensity, we decompose the "other skills" category (i.e., all skills except social and cognitive) into sub-categories. We build on Guadalupe et al. (2014), who identify distinct functions within organizations. Drawing on their classification, we divide the other skills bucket into five functional skill sub-categories: (1) customer service, (2) manufacturing operations, (3) research and development, (4) information technology, and (5) functional administrative skills, comprising finance, human resources, and legal skills. We then further classify the non-functional skills related to business, administration, and policy into two buckets: general (e.g., business, administration, or policy related skills) and industry-specific (e.g., agriculture, education, or healthcare related skills). This results in seven skill sub-categories. We manually map Lightcast skills onto these sub-categories using their skill labels, and Appendix F provides illustrative examples of the skills in each sub-category. Figure 6(a) reports the results of this decomposition.

Figure 6(a) shows that the observed increase within the other skills is driven primarily by an increase in functional administrative skills ($\beta = 0.026$, $p = 0.011$), suggesting two possibilities. First, GenAI roles within finance, human resources, or legal occupations may be expanding their existing functional skill requirements within their own corresponding functions (e.g., finance roles requiring additional finance skills), suggesting within-function skill *deepening*. Second, GenAI roles outside these functions may be expanding to include these skills, suggesting cross-functional skill *extension*. To determine where the effects are concentrated, we next perform a role-level analysis in Section 6.2.

## 6.2. Roles driving the main results

To further examine whether the *skill reweighting* effect is predominantly *skill deepening* of functional administrative skills within finance, human resources, or legal occupations or *skill extension* in other occupations, we partition our sample into these two occupation groups based on their six-digit O*NET-SOC occupation. Appendix F provides the SOCs designated as finance, human resources, and legal occupations. All remaining SOCs have been designated as other occupations. We then re-estimate changes in functional administrative skills separately for these two groups.

Figure 6(b) shows that the increase in functional administrative skills is concentrated in GenAI roles outside finance, HR, and legal occupations ($\beta = 0.027$, $p = 0.017$). By contrast, GenAI roles within these administrative functions exhibit no change ($\beta = 0.017$, $p = 0.138$). These results suggest that the observed increase in functional administrative skills in GenAI roles reflects an *extension* of finance, HR, and legal skill requirements into GenAI roles located outside those functions.

Although this pattern is consistent with *skill extension*, it may still reflect two distinct underlying requirements. First, it could imply a requirement for these roles to "self-serve" using GenAI instead of relying on colleagues in HR, finance, or legal functions. Alternatively, it could also be the case that GenAI role incumbents now need to coordinate more intensively with colleagues in these administrative functions, and the greater demand for these functional skills signals an attempt to build the requisite common ground.

We test for these possibilities by leveraging a partition into managerial versus non-managerial occupations. Since managerial work is more coordination intensive than non-managerial work (Goldberg & Puranam, 2026; Mintzberg, 1973; Zhang, 2023), effects concentrated in managerial occupations would suggest stronger support for the increased coordination requirements argument. Instead, effects concentrated in non-managerial occupations, known to be less oriented towards coordination than managerial roles, would suggest stronger support for the self-service hypothesis.

For this, we designate GenAI job postings in the O*NET SOC "11-management occupations" category as managerial and all others as non-managerial. Figure 6(b) shows that the reduction in relative demand for social skills in GenAI roles following the introduction of ChatGPT is concentrated in non-managerial roles, driven by the increase in their functional administrative skill requirements ($\beta = 0.030, p = 0.009$). Managerial GenAI roles, on the other hand, exhibit no change ($\beta = 0.005$, $p = 0.689$). These patterns suggest that the self-service interpretation is more plausible as a mechanism driving the results.

In sum, these supplementary analyses show that skill reweighting is driven primarily by skill extension of finance, HR, and legal skill requirements among non-managerial GenAI roles outside those functions. This pattern is most consistent with GenAI enabling workers to self-serve on administrative tasks rather than increasing cross-functional coordination demands.

--- INSERT FIGURE 6 ABOUT HERE ---

## 7. DISCUSSION AND CONCLUSION

### 7.1. Discussion

GenAI's capabilities in cognitive tasks have led some (Hernholm, 2025; Madgavkar et al., 2024; PwC, 2026; Raman & Flynn, 2024) to predict that as the technology takes over cognitive work, workers will focus more on interpersonal collaboration, amplifying the demand for social skills. We develop a skill-centric framework in this paper to assess this prediction, extending prior research on technology-task relationships (e.g., Acemoglu et al., 2025; Acemoglu & Restrepo, 2019). A key implication of the framework is that a technology can raise or lower the relative importance of a skill depending on whether it changes task composition, the proficiency of skill application, or whether it affects other skills in the portfolio. The framework, therefore, helps us analyze how roles evolve as new technologies are incorporated into them, complementing prior work focused on shifts in employment volume or task bundles (e.g., Dell'Acqua et al., 2023; Handa et al., 2025; Hoffmann et al., 2024; Otis et al., 2023).

Our results provide an empirical pattern that contrasts with the dominant narrative about social skills. Analyzing job postings from 595 early GenAI-adopting firms and applying the framework, we find no evidence of social skill amplification. Instead, we find evidence for re-weighting away from social skills, i.e., GenAI appears to shift tasks to rely on skills other than social ones, therefore depressing the relative importance of social skills. On the other hand, we find no changes in the relative demand for cognitive skills. Absolute skill count analysis shows no significant change in the count of social skills but an increase in the count of other skills, specifically functional administrative skills, i.e., those tied to finance, human resources, and legal functions, implying that social skills become relatively less emphasized because the role's skill portfolio expands.

Additional analyses show that this rise in functional administrative skills is concentrated outside the relevant functions, indicating that the reweighting reflects cross-functional skill extension, is strongest in non-managerial roles which are less coordination-oriented than managerial ones, and is not accompanied by greater demand for social skills. Together, we read this as evidence that GenAI roles increasingly self-serve tasks once handled by administrative functions, rather than intensify their coordination with those functions.

These findings contribute to research on the impact of technology on the nature of work (e.g., Acemoglu & Autor, 2011; Adner et al., 2019; Bresnahan, 2011) in three ways. First, conceptually, we offer a framework that shifts from tasks to skills and clarifies how task-level mechanisms may translate into either increases or decreases in relative skill demand within roles. Second, empirically, we provide evidence on how GenAI adoption is associated with changes in stated skill demand, complementing prior research that focuses on role-level or task-level outcomes (e.g., Dell'Acqua et al., 2023; Handa et al., 2025; Hoffmann et al., 2024; Otis et al., 2023). Third, our results provide an empirical pattern that contrasts with the dominant narrative about social skills. In early adopting public firms, the observed pattern instead suggests that GenAI enables the accumulation of additional cross-functional administrative skills within non-managerial roles without a corresponding increase in the emphasis on social skills. Taken together, these results suggest that GenAI's early labor-market impact may lie less in intensifying coordination and more in enabling workers to self-service tasks once managed by those administrative functions.

The results also offer practical insights for workers and firms preparing to adjust to GenAI-augmented organizations. If the observed pattern – that social skills become relatively less emphasized and that work plausibly shifts towards self-service where it previously required coordination – extends beyond early adopters and persists in the longer run, it could have adverse implications for the future employability of individuals in these roles by reducing opportunities to exercise and develop social skills. At the same time, employers may need to reassess the extent to which they rely on social skills as a foundation for building firm-specific human capital and competitive advantage.

More broadly, our findings also speak to the growing emphasis in workforce development on social and interpersonal skills as protection against automation (Madgavkar et al., 2024; WEF, 2025). The general prescription assumes that AI will increase demand for social skills. In contrast, our results suggest that in roles adopting AI, social skill demand may be falling, particularly in non-managerial settings. The return to social skills has also grown over recent decades, partly because these skills were increasingly required by cognitively intensive roles (Deming, 2017; Weinberger, 2014). If GenAI dilutes the skill requirements, workers' wage returns on these skills may fall, diminishing their opportunities to develop them, and calling into question the prevailing policy emphasis on social skills as the durable complement to technology.

### 7.2. Limitations

We acknowledge several limitations to our empirical approach. First, job postings, which we use to measure role-level GenAI adoption and skill intensity, reflect employers' stated requirements rather than the full set of competencies a role might demand. Firms may omit skills and requirements they take for granted or consider bundled with other skills and requirements. As a result, both our role-level GenAI adoption and skill intensity measures may be conservative estimates across the sample. Although our job-posting-based measures of cognitive and social-skill intensity correlate strongly with O*NET ratings and our GenAI adoption metrics are consistent with prior benchmarks of GenAI adoption (e.g., Bick et al., 2024; Eloundou et al., 2023), future research using complementary data sources, such as worker surveys or time-use data, could help triangulate the results we present.

Second, we note that while job postings reflect firm-level work requirements, they may also be influenced by factors beyond internal role characteristics, e.g., industry-wide norms such as shifting HR practices or bandwagon effects among competitors, macro-economic conditions such as labor market tightness, or societal and cultural shifts, such as growing emphasis on certain soft skills or types of initiatives across sectors, which may shape how industry players frame their jobs to applicants.

Third, the setting of U.S. public firms from 2022 to 2024 allows us to capture early organizational responses to GenAI, and these patterns may evolve as the technology adoption matures. We acknowledge this limitation and interpret our findings as evidence on early adoption dynamics rather than steady-state effects.

### 7.3. Future research

Future work can build on this study in several ways. First, because our setting captures early adoption, an important next step is to examine the longer-run impact of GenAI on skill demand as firms redesign workflows and develop complementary practices. Second, lab experiments that vary access to GenAI tools or specific features can improve our understanding of the technology's impact on social skills by directly measuring how GenAI changes interaction frequency, coordination processes, and the content of interpersonal work. Third, our skill-centric framework can be extended to incorporate complementarities between skills within the same tasks, allowing researchers to examine how technology affects these complementarities over time. Fourth, and more broadly, our conceptual model can be extended along two dimensions. Beyond GenAI, it can be applied to other technologies to clarify the conditions under which technology leads to upskilling, deskilling, or reskilling within roles. Beyond the cognitive-social skill distinction we explore, it can also incorporate other skills in general, and specifically a broader array of non-cognitive attributes such as perseverance, motivation, self-control, and self-esteem, that may also shape labor-market outcomes in important ways (Cunha & Heckman, 2007; Heckman et al., 2006; Lindqvist & Vestman, 2011). The latter extension may be particularly fruitful if GenAI roles increasingly self-serve tasks once handled through coordination, and demand may be shifting towards self-management traits such as conscientiousness and self-control. Fifth, while our analysis focuses on how GenAI transforms existing roles, future research could also examine how the technology drives the creation of entirely new roles and elimination of existing ones.

### 7.4. Conclusion

In conclusion, our findings challenge the common view that GenAI will reduce the importance of cognitive skills while amplifying demand for social skills by shifting workers toward interpersonal coordination. Evidence from early GenAI-adopting U.S. public firms points to a different initial trajectory: GenAI roles add skills – especially functional administrative ones – lowering the relative emphasis on social skills while leaving cognitive-skill intensity unchanged. These patterns are concentrated in non-managerial settings, suggesting that such roles increasingly self-serve tasks previously performed by administrative functions. However, if a similar trajectory extends over time into different settings, it would imply a future that departs sharply from the prevailing expectation of rising social-skill importance.

## 8. AI USE STATEMENT

The authors used OpenAI's ChatGPT and Anthropic's Claude, across multiple model versions, to improve the clarity of written text. All AI-generated content was reviewed, verified for accuracy, and edited by the authors. The authors take full responsibility for the accuracy and integrity of all material in the manuscript.

## 10. TABLES

**Table 1.** Illustration of the many-to-many mapping between technology adoption effects on tasks (complementarity, displacement, and creation) vs. skills (deskilling/upskilling, displacement/expansion, reweighting).

| Technology adoption's effect on underlying tasks | Technology adoption's effect on skill X | Label | Example (Skill X in *italics*) |
|---|---|---|---|
| **Decrease in relative demand for skill X** | | | |
| **I. No change in tasks** | Technology lowers required proficiency at skill X. | Deskilling | Grammarly reduces the demand for clear and correct *writing skills* (same quality standard can be met with lower writing proficiency). |
| **II. Existing human task displaced** | Eliminates an existing task that required skill X. | Skill displacement | PowerBI automates analytical work, decreasing the demand for *analytical skills*. |
| **III. New human task created** | Creates a new task that to leverage other skills than X. | Skill reweighting | Spreadsheets create electronic data-entry tasks, reducing the relative demand for *accounting skills* (denominator increases). |
| **Increase in relative demand for skill X** | | | |
| **I. No change in tasks** | Technology raises required proficiency at skill X | Upskilling | Generative-AI coding assistants raise demand for *architectural code-evaluation skills* as human must judge and integrate code |
| **II. New human task created** | Creates a new task that requires skill X. | Skill expansion | The rollout of AI copilots creates prompt-engineering tasks, increasing the demand for *prompt-design skills.* |
| **III. Existing human task displaced** | Eliminates an existing task that required other skills than X. | Skill reweighting | ATMs automate cash-handling work, increasing the relative importance of customer-service skills (denominator falls). |

*Note.* A null change in skill demand is also possible if multiple mechanisms co-occur and offset one another.

**Table 2.** Variable description, summary statistics, and pairwise correlation.

| Variable | Description | Mean | Std Dev. | (1) | (2) | (3) | (4) | (5) | (6) | (7) |
|---|---|---|---|---|---|---|---|---|---|---|
| (1) Cognitive Skill Intensity | Count of cognitive skills as a percentage of total count, averaged at the firm-role-month level | 2.938 | 4.225 | 1.000 | | | | | | |
| (2) Social Skill Intensity | Count of social skills as a percentage of total count, averaged at the firm-role-month level | 3.272 | 4.132 | 0.072 | 1.000 | | | | | |
| (3) Cognitive Skills Count | IHS transformed[1] count of skills corresponding to the keywords[2]: problem solving, research, analytical, critical thinking, math, statistics, averaged at the firm-role-month level | 0.455 | 0.553 | 0.817 | 0.112 | 1.000 | | | | |
| (4) Social Skills Count | IHS transformed[1] count of skills corresponding to the following keywords[2]: communication, teamwork, collaboration, negotiation, presentation, averaged at the firm-role-month level | 0.498 | 0.543 | 0.136 | 0.818 | 0.314 | 1.000 | | | |
| (5) Other Skills Count | IHS transformed[1] count of non-cognitive non-social other skills, averaged at the firm-role-month level | 3.309 | 0.691 | 0.111 | 0.083 | 0.413 | 0.407 | 1.000 | | |
| (6) Manager | A dummy variable with value 1 for managerial roles, i.e., those with two-digit SOC "11-Managerial Occupations", and 0 otherwise | 0.141 | 0.348 | 0.057 | 0.091 | 0.105 | 0.162 | 0.130 | 1.000 | |
| (7) Functional Admin Role | A dummy variable with value 1 for functional administrative roles[3], and 0 otherwise | 0.068 | 0.251 | 0.075 | 0.063 | 0.105 | 0.088 | 0.067 | 0.133 | 1.000 |

Note: 1. IHS transform = Inverse hyperbolic sine transform, 2. The keywords for cognitive and social skills are obtained from Deming & Kahn (2018), 3. Appendix Table F3 provides the O*NET six-digit SOCs designated as functional administrative roles.

**Table 3.** Main results.

| | (1) Cognitive Skills Intensity | (2) Social Skills Intensity | (3) Cognitive Skills Count | (4) Social Skills Count | (5) Other Skills Count |
|---|---|---|---|---|---|
| GenAI Role × Post ChatGPT Launch | -0.050 | -0.111 | -0.005 | -0.008 | 0.024 |
| | (0.122) | (0.040) | (0.015) | (0.009) | (0.011) |
| | [0.681] | [0.006] | [0.723] | [0.391] | [0.031] |
| Firm-Role FE | Y | Y | Y | Y | Y |
| Month FE | Y | Y | Y | Y | Y |
| Observations | 1,709,731 | 1,709,731 | 1,709,731 | 1,709,731 | 1,709,731 |
| Num firms | 595 | 595 | 595 | 595 | 595 |
| Num roles | 41,852 | 41,852 | 41,852 | 41,852 | 41,852 |
| Adj R-square | 0.652 | 0.663 | 0.707 | 0.695 | 0.780 |
| Mean DV | 2.938 | 3.272 | 0.455 | 0.498 | 3.309 |

Note: Models 1 and 2 report post-ChatGPT launch (November 2022) changes in *skill intensity* for cognitive and social skill categories, while models 3-5 report changes in *Skill Count* for cognitive, social, and other skill categories. *Skill intensity* is calculated as the count of skills within the relevant category, expressed as a percentage of total skill count for a job posting, and then averaged at the firm-role-month level. *Skill count* is calculated as the inverse hyperbolic sine transform of the raw skill count and then averaged at the firm-role-month level. The variable *GenAI role* equals 1 if the text of at least one job posting within the focal role after November 2022 includes GenAI keywords. Other roles within the same six-digit SOC occupation are used as the counterfactual group. The table reports coefficients on the interaction term *GenAI role × post ChatGPT* from a difference-in-differences OLS regression estimated at the firm-role-month level, with job-posting count used as weights, firm-role fixed effects, and month fixed effects. Standard errors, clustered at the firm and role levels, are reported in parentheses. p-values are reported in square brackets below the standard errors.

## 11. FIGURES

**Figure 1.** Descriptive statistics: Monthly distribution of advertised GenAI job postings.

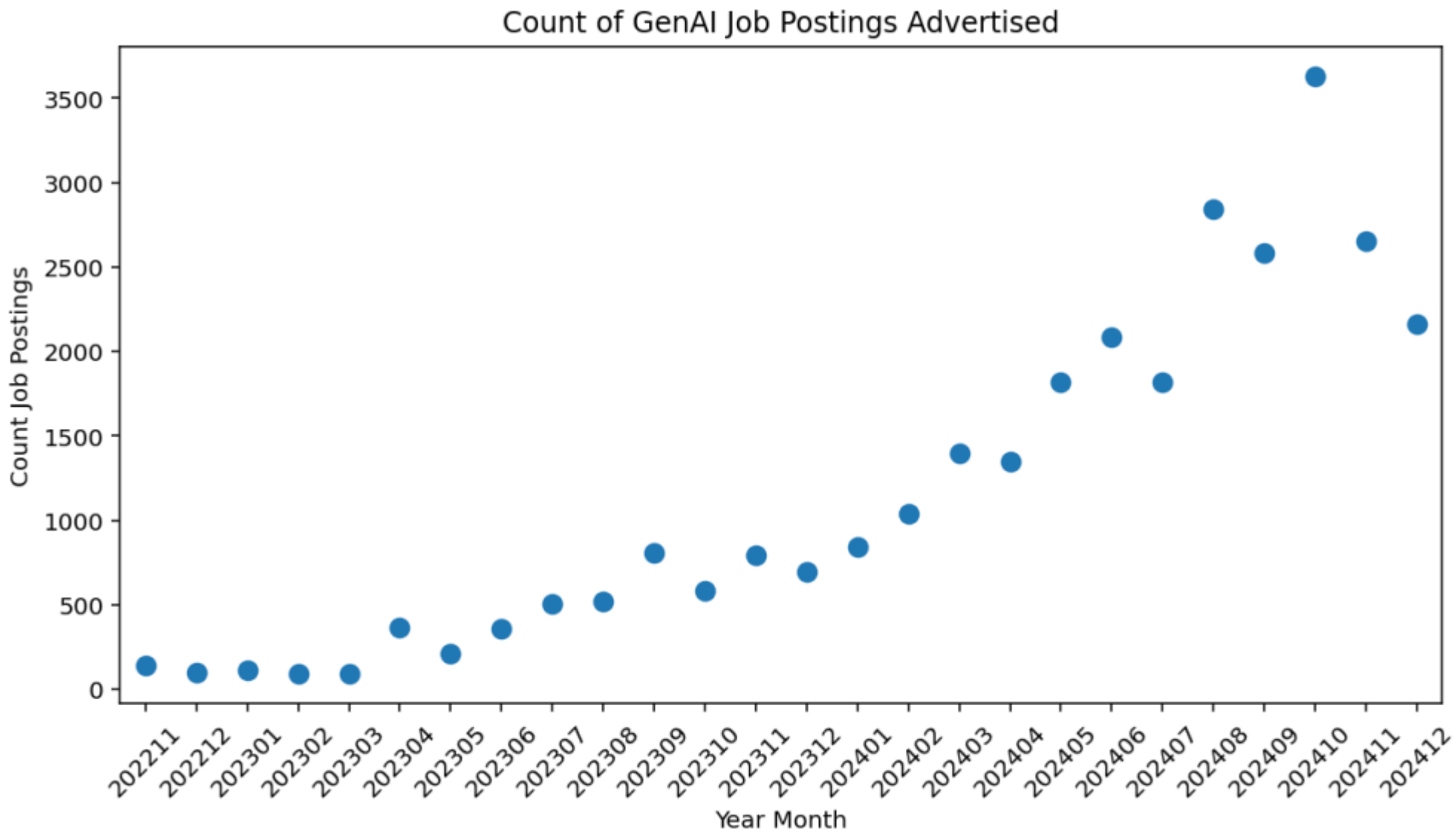

**Figure 2.** Descriptive statistics: GenAI job postings by occupation and industry.

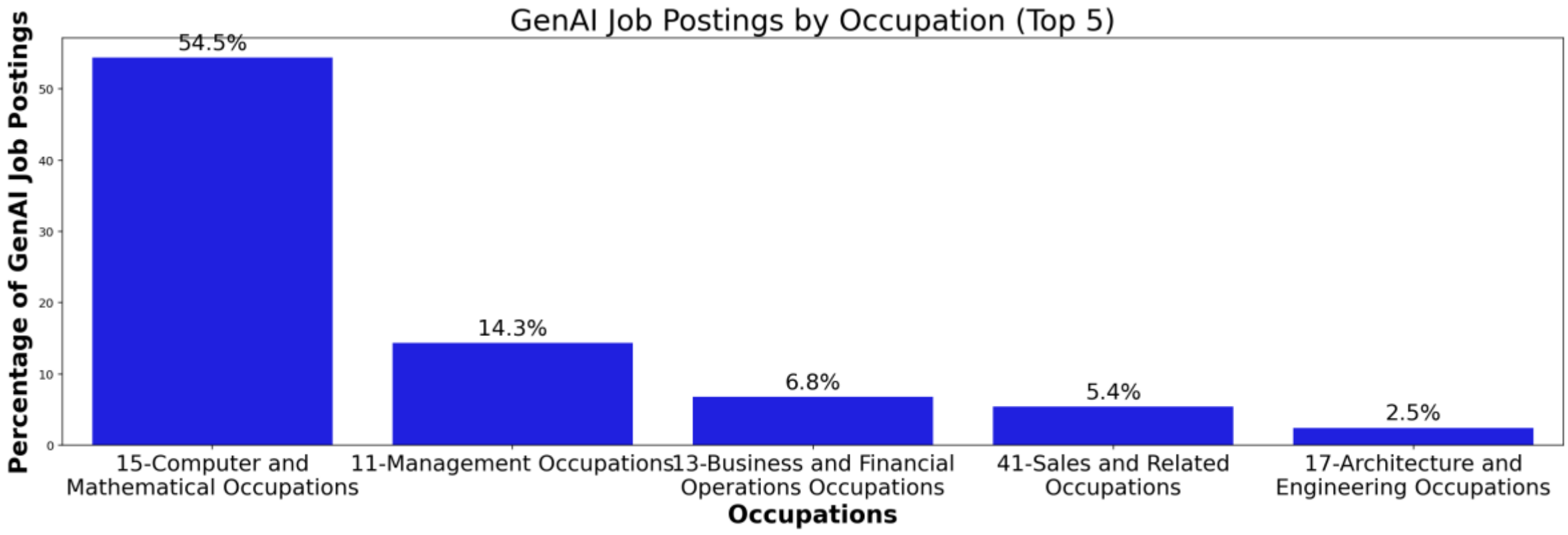

Figure 2 (a): GenAI job postings by Occupation (Two-digit SOC code)

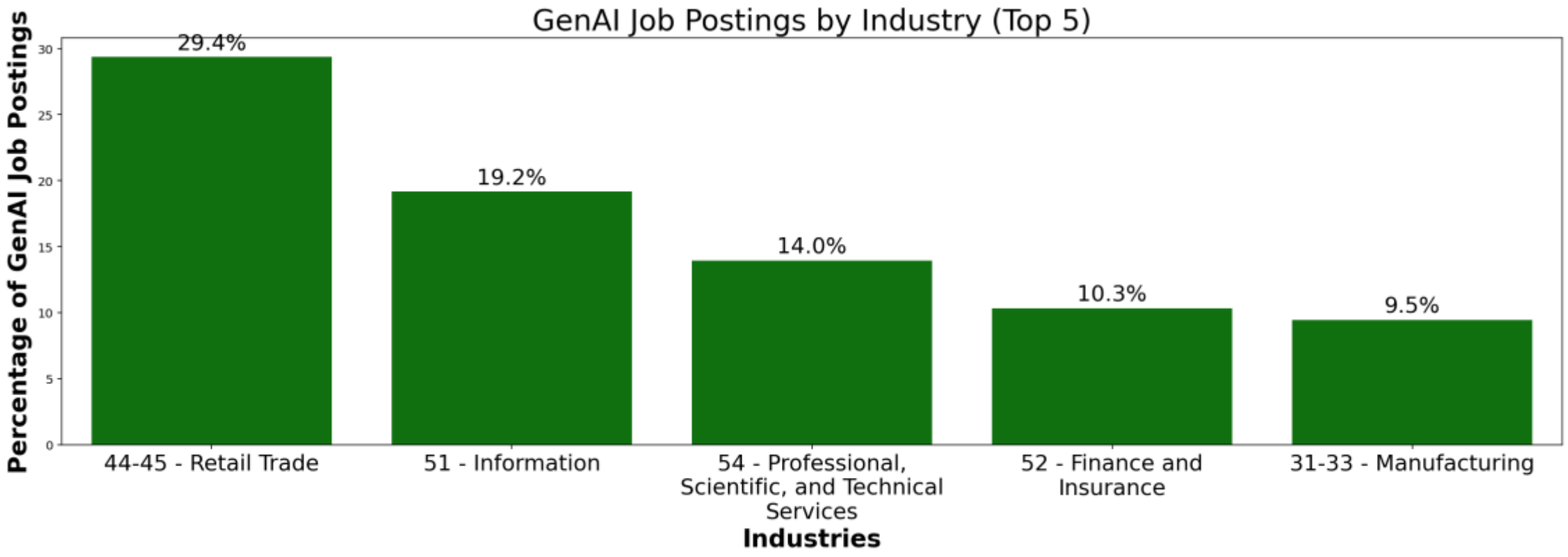

Figure 2 (b): GenAI job postings by Industry (Two-digit NAICS code)

**Figure 3.** Cross-sectional analyses: Comparing *skill intensity* for GenAI roles with other roles before ChatGPT launch.

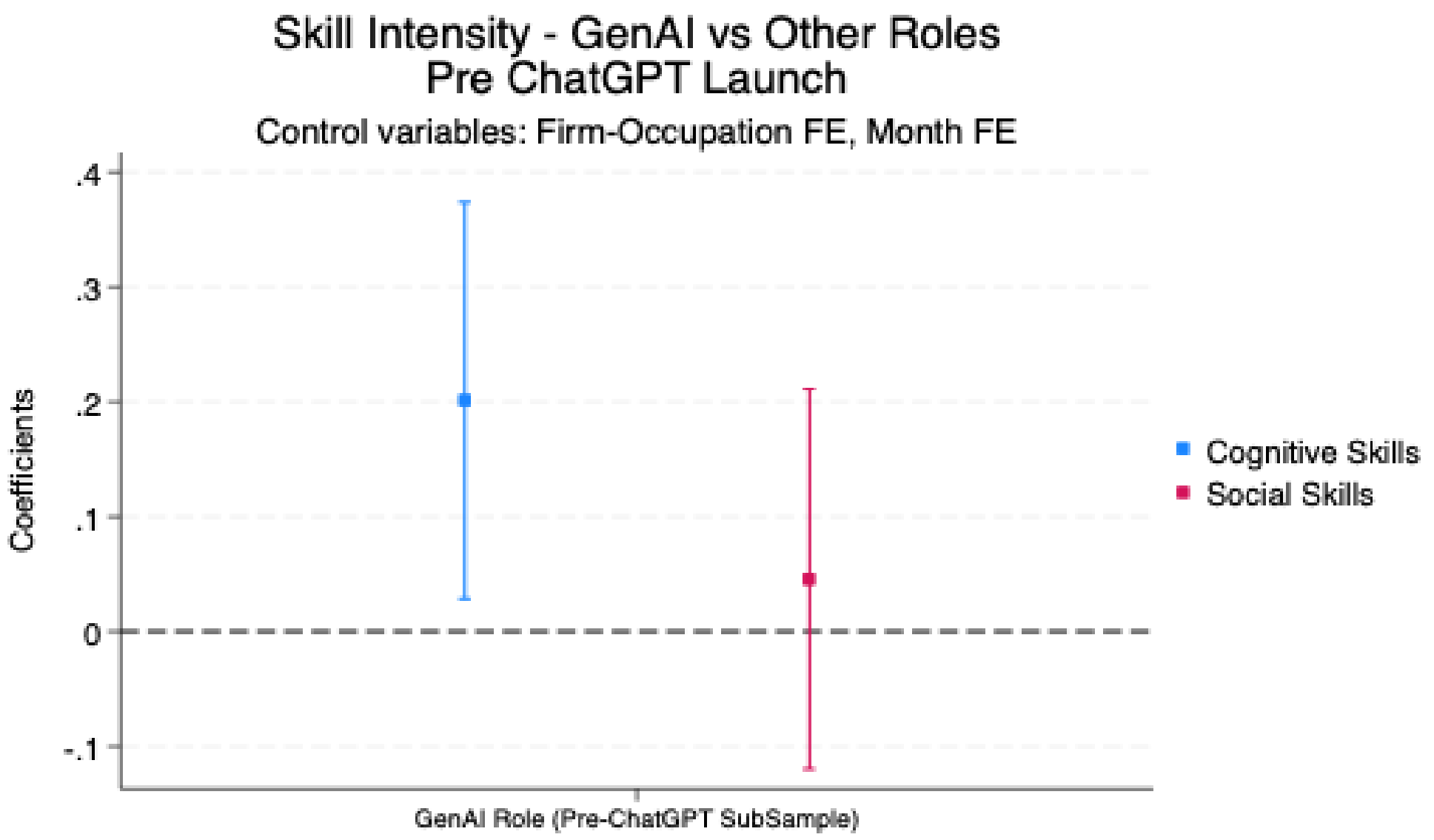

Note: The figure above plots the differences in *skill intensity* for cognitive and social skill categories before the launch of ChatGPT (between January and October 2022), comparing *GenAI roles* with other roles within the same occupation (six-digit SOC level). *Skill intensity* is calculated as the count of skills within the relevant category, expressed as a percentage of total skill count for a job posting, and then averaged at the firm-role-month level. Coefficients on the *GenAI role* indicator from OLS regressions estimated at the firm-role-month level, with job-posting count used as weights, firm-occupation fixed effects, and month fixed effects have been plotted. Standard errors are clustered at the firm and occupation levels. The figure reports 95% confidence intervals.

**Figure 4.** Difference-in-differences analyses: Comparing *skill intensity* for GenAI roles, pre and post ChatGPT launch, adjusting for changes within other roles.

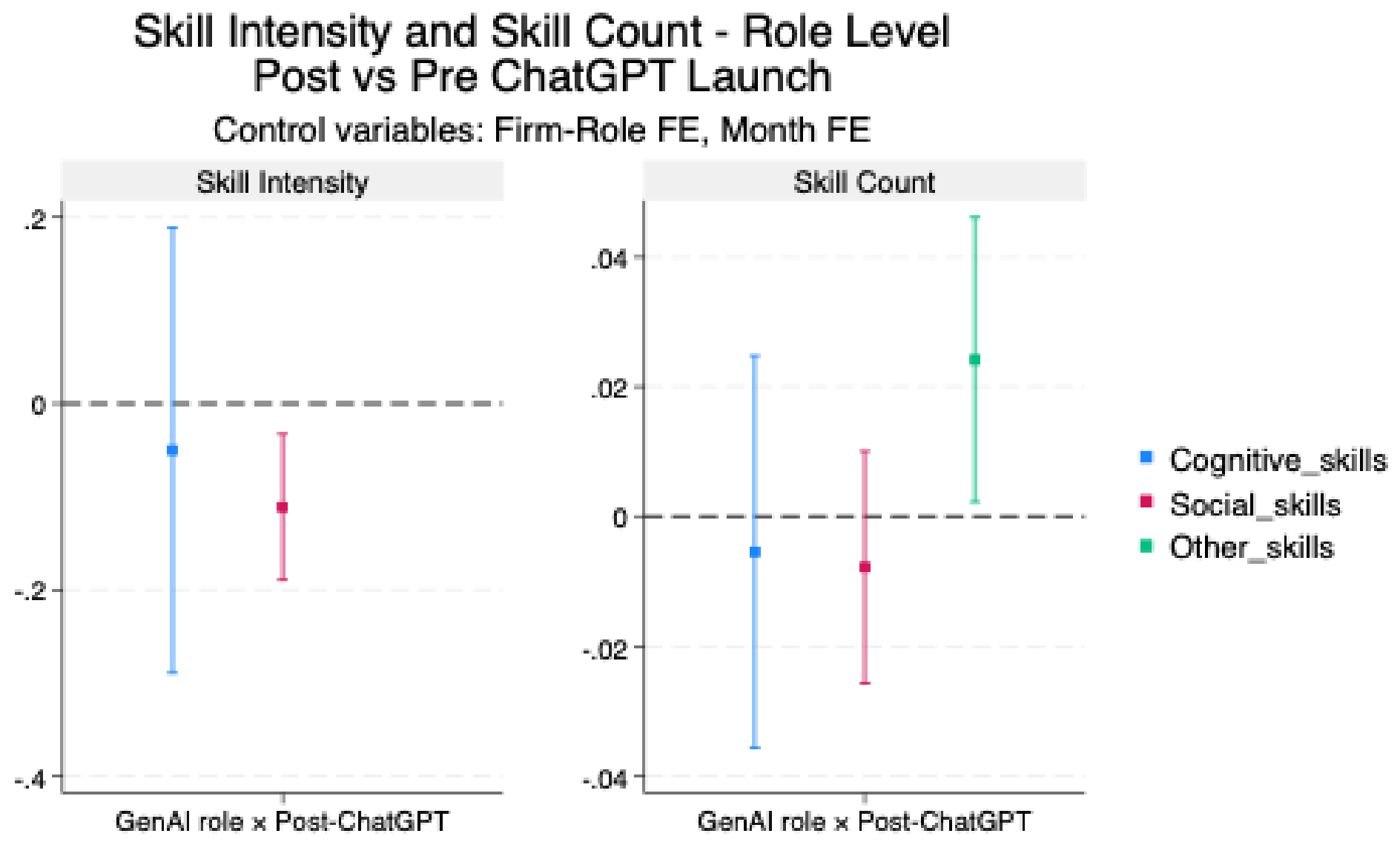

Note: The figure above plots post-ChatGPT launch changes in *skill intensity* for the cognitive and social skill categories, and changes in *skill count* for cognitive, social, and other skill categories among GenAI roles. *Skill intensity* is calculated as the count of skills within the relevant category, expressed as a percentage of total skill count for a job posting, and then averaged at the firm-role-month level. *Skill count* is calculated as the inverse hyperbolic sine transform of the raw skill count and then averaged at the firm-role-month level. The variable *GenAI role* equals 1 if the text of at least one job posting within the focal role after November 2022 includes GenAI keywords. Other roles within the same six-digit SOC occupation are used as the counterfactual group. Coefficients on the interaction term *GenAI role × post ChatGPT* from a difference-in-differences OLS regression estimated at the firm-role-month level, with job-posting count used as weights, firm-role fixed effects, and month fixed effects have been plotted. Standard errors are clustered at the firm and role levels. The figure reports 95% confidence intervals.

**Figure 5:** Time-series estimates: Month-over-month comparison of changes in *skill intensity* and *skill count* for GenAI roles, pre and post ChatGPT launch, adjusting for changes within other roles.

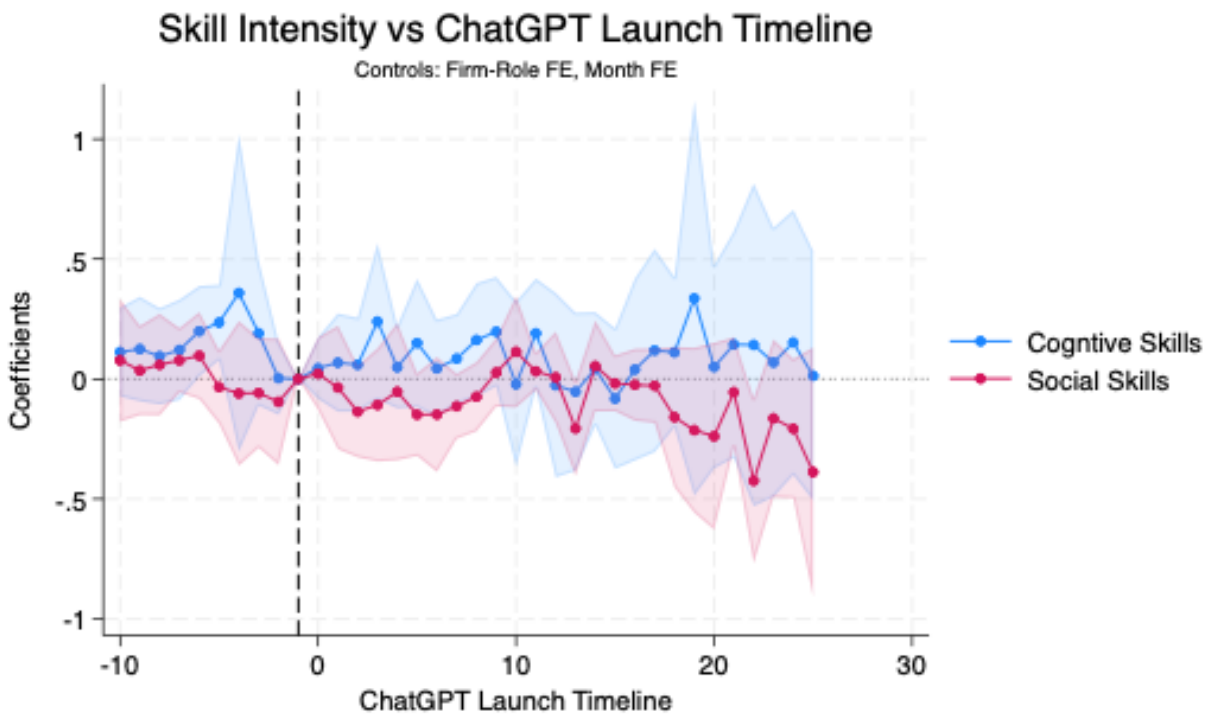

**Figure 5(a):** Skill intensity time series estimates

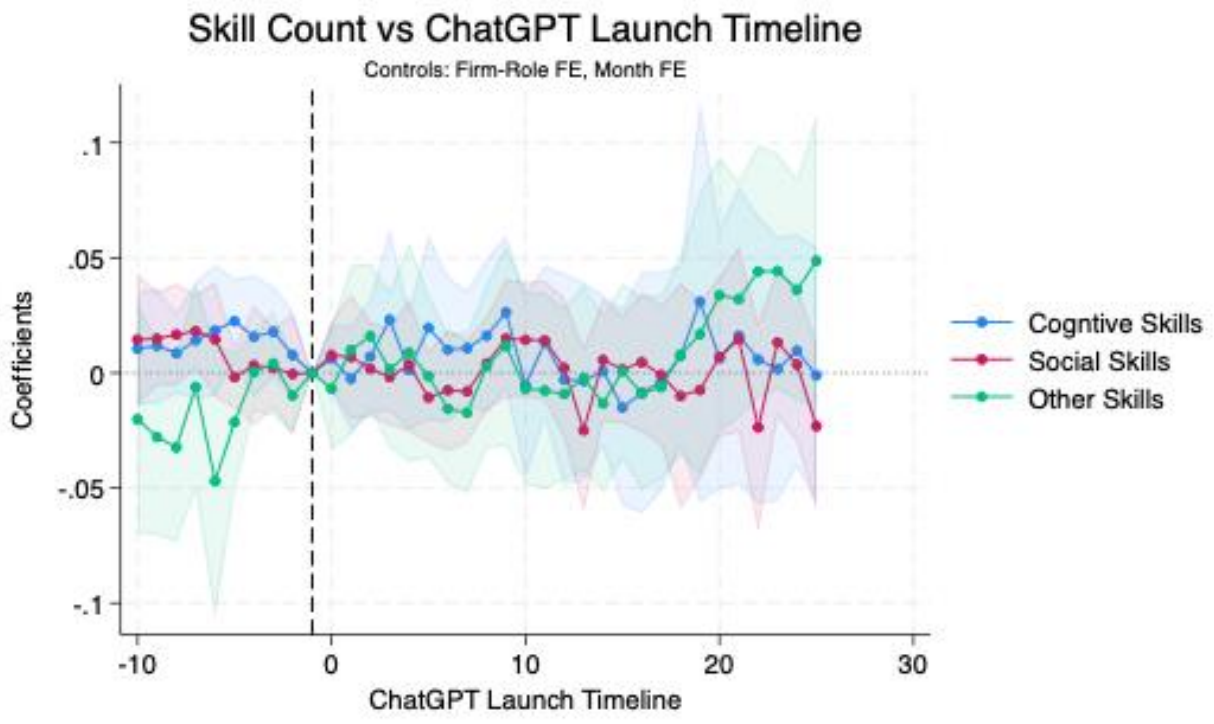

**Figure 5(b):** Skill count time series estimates

Note: The figures above plot time-series estimates of changes in *skill intensity* for the cognitive and social skill categories (Figure 5a) and *skill count* for cognitive, social, and other skill categories (Figure 5b) among GenAI roles. *Skill intensity* is calculated as the count of skills within the relevant category, expressed as a percentage of total skill count for a job posting, and then averaged at the firm-role-month level. *Skill count* is calculated as the inverse hyperbolic sine transform of the raw skill count and then averaged at the firm-role-month level. The variable *GenAI role* equals 1 if the text of at least one job posting within the focal role after November 2022 includes GenAI keywords. Other roles within the same six-digit SOC occupation are used as the counterfactual group. Coefficients on the interaction terms *GenAI role* × *month* from event-study OLS regressions estimated at the firm-role-month level have been plotted, with October 2022 (the month before ChatGPT's launch) as the omitted base month. The specifications are otherwise similar to those reported in Figures 4, with job-posting count used as weights, firm-role fixed effects, and month fixed effects. Standard errors are clustered at the firm and role levels. The figures report 95% confidence intervals.

**Figure 6.** Supplementary analyses.

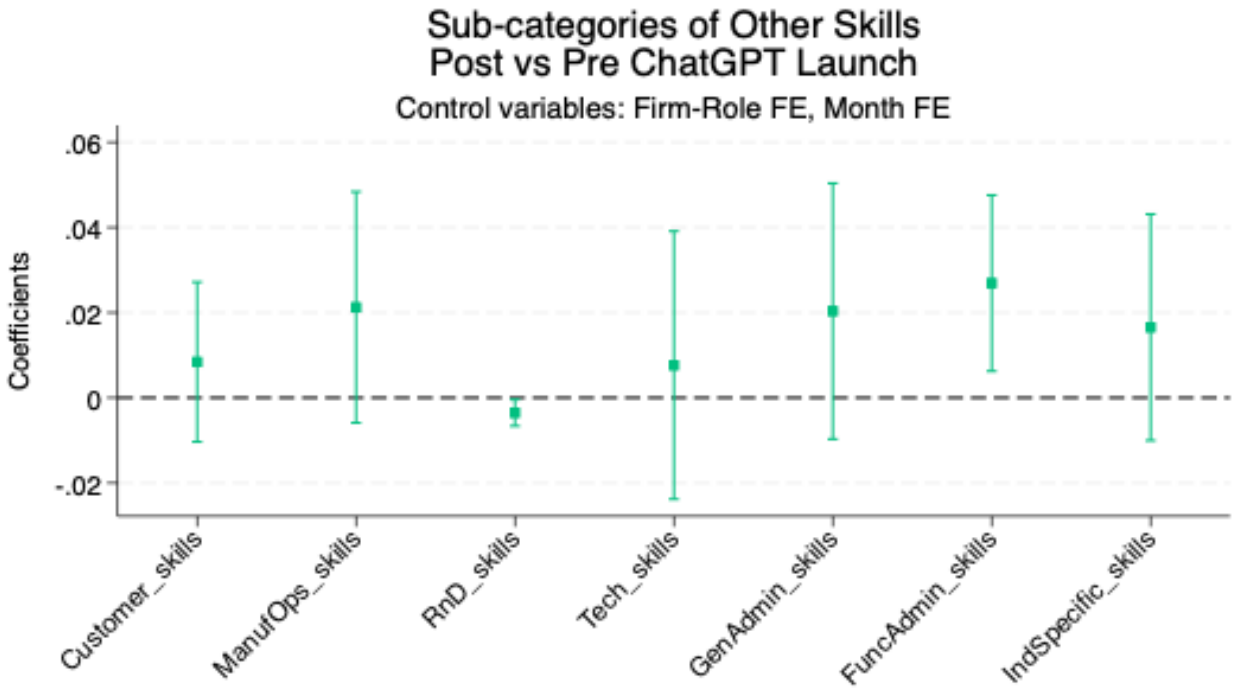

**Figure 6(a):** Sub-categories of other skills

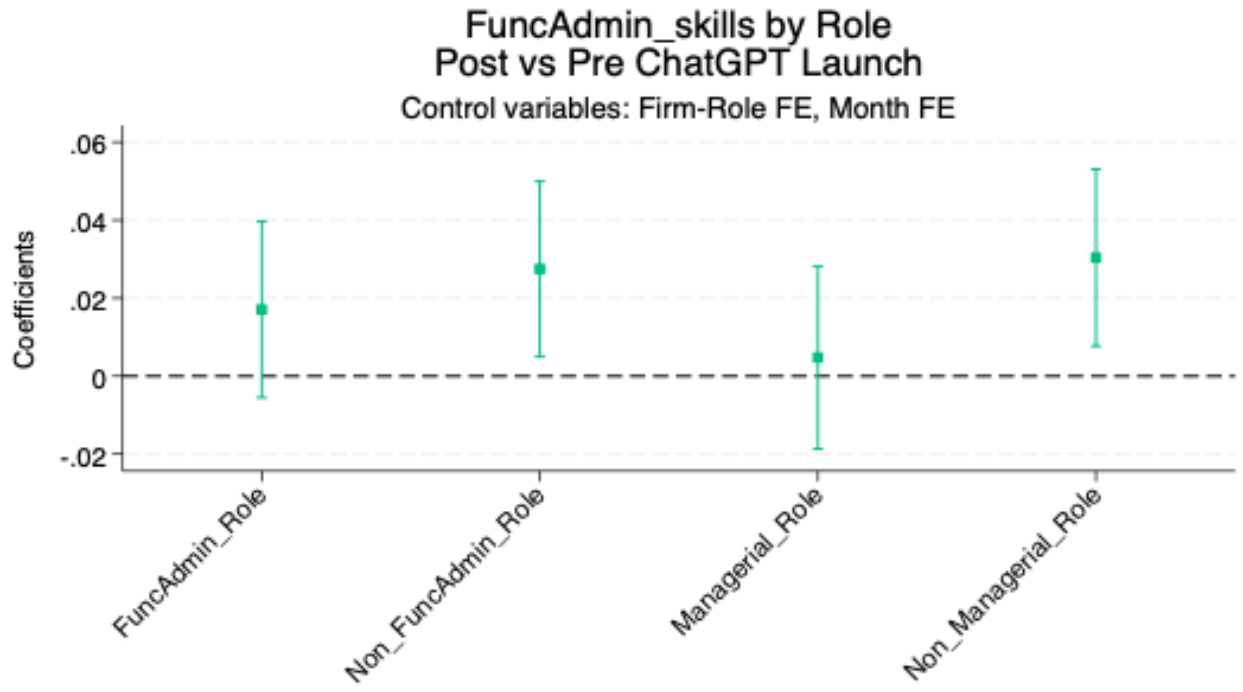

**Figure 6(b):** Functional admin skills by role

Note: Figure 6(a) above plots post-ChatGPT launch changes in different sub-categories of *other skill count* among GenAI roles, relative to the pre-launch period. Each sub-category of *other skill count* is measured as the inverse hyperbolic sine transform of the raw count of skills in that subset, averaged at the firm-role-month level. The other skills, i.e., non-social and non-cognitive skills, are grouped into the following sub-categories: customer service skills (*customer_skills*), manufacturing operations skills (*manufops_skills*), research and development skills (*rnd_skills*), information technology skills (*tech_skills*), general business and administrative skills (*genadmin_skills*), finance, human resources, legal functional administrative skills (*funcadmin_skills*), and other industry-specific skills (*indspecific_skills*), based on Lightcast skill labels. In addition, Figure 6(b) above splits changes in functional administrative skills by (1) whether the GenAI role belongs to a functional administrative occupation (i.e., those in finance, HR, or legal occupation, labeled in the figure as *funcadmin_role*), or to any other occupation (labeled in the figure as *non_funcadmin_role*), and (2) whether the GenAI role belongs to a managerial (two-digit SOC "11-Managerial Occupations", labeled in the figure as *managerial_role*) or a non-managerial occupation (all others 2-digit SOC codes, labeled in the figure as *non_managerial_role*). Other roles within the same six-digit SOC occupation are used as the counterfactual group. Coefficients on the interaction term *GenAI role* × *post ChatGPT* from a difference-in-differences OLS regression estimated at the firm-role-month level, with job-posting count used as weights, firm-role fixed effects, and month fixed effects have been plotted. Standard errors are clustered at the firm and role levels. The figure reports 95% confidence intervals.

# ONLINE APPENDIX

## Appendix A: Lightcast Granular Roles

The 6,961,472 job postings advertised between 2022 and 2024 by GenAI-adopting U.S. public firms are classified into 214 O*NET Standard Occupational Classification (SOC) codes at the six-digit level. The job postings are also classified by Lightcast into 41,852 granular roles. Table A1 provides a sample of the mapping between Lightcast roles and O*NET SOC codes, showing the five most common Lightcast roles for each of the ten most common SOC codes in our data.

**Table A1.** Lightcast role to O*NET SOC mapping (for top five SOCs by job ad frequency).

| Lightcast Role | Lightcast Role Description | O*NET SOC | O*NET SOC Description |
|---|---|---|---|
| ETBCF5DD72BB5BC69C | Shift Supervisors | 11-1021 | General and Operations Managers |
| ET29F7F5758B87D9D2 | Operations Supervisors | 11-1021 | General and Operations Managers |
| ET07C3AE882A44AD54 | Sales and Operations Management Trainees | 11-1021 | General and Operations Managers |
| ET06B4D0E32816A4BF | Retail Operations Supervisors | 11-1021 | General and Operations Managers |
| ET7D1102D3A3085650 | Freight Operations Supervisors | 11-1021 | General and Operations Managers |
| ETAA6FBD39376F5BBC | Product Managers | 11-2021 | Marketing Managers |
| ETE8D6674341494CFC | Principal Product Managers | 11-2021 | Marketing Managers |
| ETDF74F02BE8537D0E | Marketing Managers | 11-2021 | Marketing Managers |
| ET15F6A253A4D313D5 | Digital Product Managers | 11-2021 | Marketing Managers |
| ET535279A2C7F23FCA | Technical Product Managers | 11-2021 | Marketing Managers |
| ET6850661D6AE5FA86 | Software Engineers | 15-1252 | Software Developers |
| ETC6F5E57496B18612 | Software Development Engineers | 15-1252 | Software Developers |
| ETFDFAE7798F611D68 | Principal Software Engineers | 15-1252 | Software Developers |
| ETC5BFB3CF8BC43347 | DevOps Engineers | 15-1252 | Software Developers |
| ET60D617B83DCEBBE6 | Solutions Architects | 15-1252 | Software Developers |
| ET0F6BD40248EB82F1 | Systems Engineers | 15-1299 | Computer Occupations, All Other |
| ETEF13D4DE7CDB43D4 | Principal Systems Engineers | 15-1299 | Computer Occupations, All Other |
| ETA8AE620FABF449EB | Principals | 15-1299 | Computer Occupations, All Other |
| ET44D2E23107A6EBF0 | Scrum Masters | 15-1299 | Computer Occupations, All Other |
| ETB480C279F0423146 | Product Owners | 15-1299 | Computer Occupations, All Other |
| ET3B93055220D592C8 | Data Scientists | 15-2051 | Data Scientists |
| ET3037E0C947A02404 | Data Analysts | 15-2051 | Data Scientists |
| ETE163411CFD44DA55 | Business Intelligence Engineers | 15-2051 | Data Scientists |
| ET5AA72D0E18D0EFE5 | Enterprise Architects | 15-2051 | Data Scientists |
| ETA7B9345E537B7C0A | SAP Interns | 15-2051 | Data Scientists |
| ETFDE2EC77DDCB5B2C | Operations Managers | 41-1011 | First-Line Supervisors of Retail Sales Workers |
| ET7BAD3A7D1C44B7EF | Assistant Store Managers | 41-1011 | First-Line Supervisors of Retail Sales Workers |

| Lightcast Role | Lightcast Role Description | O*NET SOC | O*NET SOC Description |
|---|---|---|---|
| ET901B13CE023A304D | Store Managers | 41-1011 | First-Line Supervisors of Retail Sales Workers |
| ET64A8FA61331F811A | Retail Store Managers | 41-1011 | First-Line Supervisors of Retail Sales Workers |
| ET1BC5D2A9C9F45F28 | Key Holders | 41-1011 | First-Line Supervisors of Retail Sales Workers |
| ET0000000000000000 | Unclassified | 41-2031 | Retail Salespersons |
| ET95AFA4F018676005 | Retail Sales Associates | 41-2031 | Retail Salespersons |
| ET80ACB6F6DCD02C9E | Sales Associates | 41-2031 | Retail Salespersons |
| ET651B3CFA1188BFE8 | Store Associates | 41-2031 | Retail Salespersons |
| ET968AF5BB48078950 | Seasonal Retail Associates | 41-2031 | Retail Salespersons |
| ET3155900D80932338 | Account Managers | 41-4012 | Sales Representatives, Wholesale and Manufacturing, Except Technical and Scientific Products |
| ET36F5867207489834 | Account Executives | 41-4012 | Sales Representatives, Wholesale and Manufacturing, Except Technical and Scientific Products |
| ET8A3B22D8551CD102 | Sales Representatives | 41-4012 | Sales Representatives, Wholesale and Manufacturing, Except Technical and Scientific Products |
| ET93AFDEFE5910651F | Enterprise Account Executives | 41-4012 | Sales Representatives, Wholesale and Manufacturing, Except Technical and Scientific Products |
| ET9D7EB9BEE816B6A8 | First Aid and Safety Service Sales Representatives | 41-4012 | Sales Representatives, Wholesale and Manufacturing, Except Technical and Scientific Products |
| ET2D1EA6A414179033 | Customer Service Representatives | 43-4051 | Customer Service Representatives |
| ET57B68CC2998B7D11 | Guest Experience Leaders | 43-4051 | Customer Service Representatives |
| ETC1F481A76BD88C5D | Bilingual Spanish Customer Service Representatives | 43-4051 | Customer Service Representatives |
| ET8E5F1B9D18DECBE2 | Customer Support Specialists | 43-4051 | Customer Service Representatives |
| ET450CA414AF18B1DD | Customer Service Associates | 43-4051 | Customer Service Representatives |
| ETB90D3BA8E6B1B682 | Warehouse Package Handlers | 53-7062 | Laborers and Freight, Stock, and Material Movers, Hand |
| ETA9D4A0FE5190F03C | Warehouse Workers | 53-7062 | Laborers and Freight, Stock, and Material Movers, Hand |
| ET3A79AE164A6C67C5 | Warehouse Associates | 53-7062 | Laborers and Freight, Stock, and Material Movers, Hand |
| ET7771B799528068F7 | Warehouse Production Workers | 53-7062 | Laborers and Freight, Stock, and Material Movers, Hand |
| ETB1A036F8525D47EE | Material Handlers | 53-7062 | Laborers and Freight, Stock, and Material Movers, Hand |

## Appendix B: Cognitive and Social Skill Intensity Measures from Lightcast

The keywords to measure cognitive and social skill intensity are derived from Deming & Kahn (2018). Table B1 reports the cognitive and social skills keywords from Deming and Kahn (2018), which we manually mapped onto the Lightcast skills (data version released in April 2025). We use the Lightcast skills to count the number of cognitive and social skills observed in posted jobs. We then calculate the relative demand intensity of these skills by dividing the number from the previous step by the total number of skills mapped to the same job posting. For analysis, we average these measures at the firm-role-month level.

**Table B1.** Lightcast skills derived from prior literature's keywords, used to identify cognitive and social skills.

| Category | Skills/Keywords |
|---|---|
| Cognitive skills | **Analytical** (Analytical Skills, Data Analysis, Business Analysis), **Critical thinking** (Critical Thinking, Decision Making), **Math** (Mathematics), **Problem solving** (Problem Solving, Creative Problem Solving), **Research** (Research, Internet Research, Market Research, Clinical Research), **Statistics** (Statistics, Statistical Analysis) |
| Social skills | **Collaboration** (Collaboration, Collaborative Communications, Coordinating), **Communication** (Effective Communication, Business Communication, Corporate Communications, Interpersonal Communications, Verbal Communication Skills, Online Communication, Persuasive Communication, Employee Communications, Lateral Communication), **Negotiation** (Negotiation, Business Negotiation, Influencing Skills), **Presentation** (Presentations, Business Presentations, Technical Presentations), **Teamwork** (Teamwork, Team Building, Team Management, Social Skills) |

Note: The keywords in **Bold** are from Deming and Kahn (2018, pS347 Table 1); those in brackets are skills from Lightcast skill taxonomy that were manually identified to exactly match or contain the keywords in bold.

To validate the measures, we use survey responses collected by the Occupational Information Network (O*NET) under the O*NET data collection program. For different occupation classes covered, O*NET requests responses from workers and occupational experts who actually work in those occupations and ask them about work attributes and job characteristics. Their responses thus reflect the true occupation characteristics. Therefore, any measure of social and cognitive skills constructed from O*NET should represent a more objective signal of the importance of social skills in organizations than job postings.[11]

For the social skill measure, we use the survey responses to the four social skill aspects described by Deming & Kahn (2018). These are: (1) 2.B.1.a: Social Perceptiveness, (2) 2.B.1.b: Coordination, (3) 2.B.1.c: Persuasion, and (4) 2.B.1.d Negotiation. For the cognitive skill measure, we use the survey responses to: (1) 2.A.1.e: Mathematics, (2) 2.A.2.a: Critical Thinking, (3) 2.B.2.i: Complex Problem Solving, and (4) 2.B.4.e Judgment and Decision

[11] As these O*NET survey scores are only updated every few years (5 to 7) and do not vary by firm, these cannot be used as the main measure in our analyses.

Making. We use the survey response database version 29.1 (released in November 2024), and leverage the level rating score by workers, which tells us "the degree, or point along a continuum, to which a particular descriptor is required or needed to perform the occupation".

Figure B1 plots the correlation between the Lightcast measure and O*NET measure when both are averaged at the six-digit O*NET SOC level. We find a high correlation of 0.537 for cognitive skills and 0.502 for social skills. These strong positive correlations with O*NET measures indicate that the job posting-based measure allows us to observe true role characteristics, to the best extent possible.

**Figure B1.** Scatterplot between O*NET survey score and Lightcast measure.

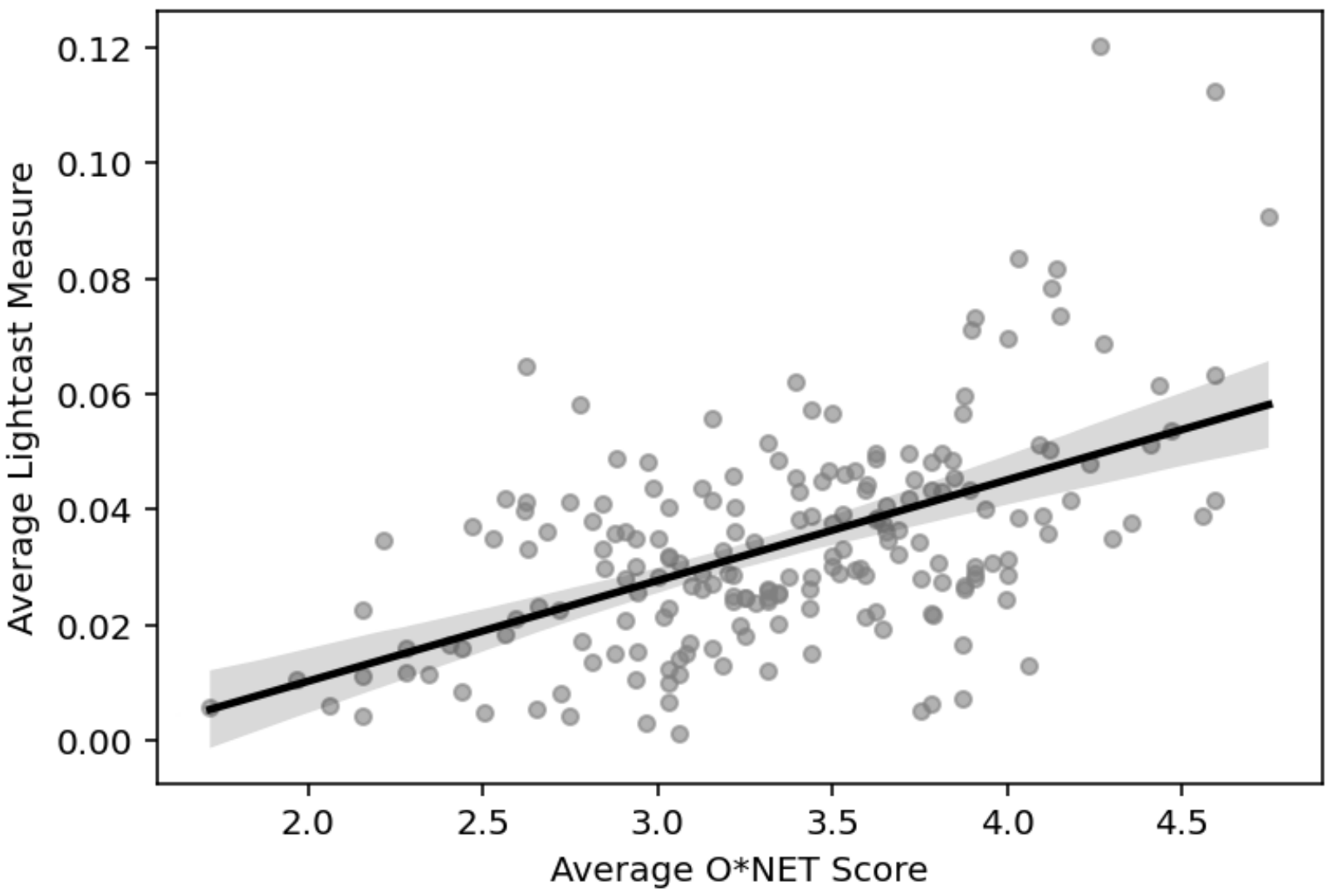

**Figure B1(a).** O*NET Cognitive Skill Level Score vs Lightcast Cognitive Skill Intensity

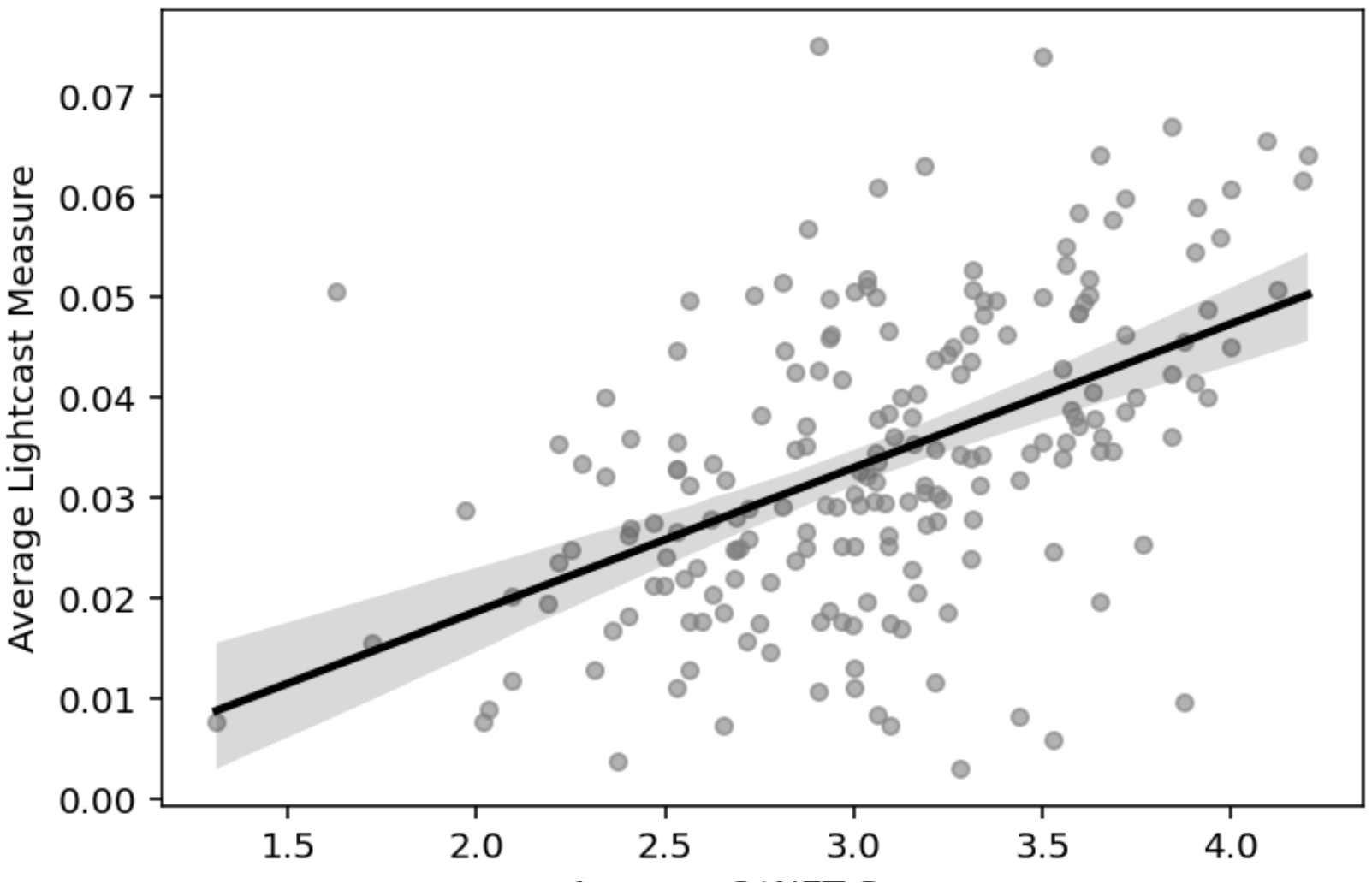

**Figure B1(b).** O*NET Social Skill Level Score vs Lightcast Social Skill Intensity

## Appendix C: GenAI Adoption Measure from Lightcast

We measure GenAI adoption for roles based on whether their job postings mention a word from a dictionary GenAI terms. This dictionary contains generic GenAI terms, named products and platforms, or GenAI-specific role descriptors. The dictionary excludes broader, non-specific terms such as “artificial intelligence” and “machine learning” that predate generative AI. Figure C1 presents snapshots of two illustrative postings identified using this measure.

To validate the GenAI adoption measure we construct from Lightcast, we compare it to data reported in Bick et al. (2024, p. 20, Fig. 8). Their study includes results from two nationally representative U.S. surveys (conducted in August and November 2024) on GenAI use at work and at home. Although the paper does not include the underlying microdata, Figure 8 reports GenAI adoption rates by occupation group (two-digit SOC). In that figure, the authors also benchmark adoption against occupation-level predicted exposure measures from Eloundou et al. (2023), which we use for additional validation tests.

To validate our operationalization, we take the following steps: for each occupation group (two-digit SOC), we first compute the share of 2024 job postings that include at least one GenAI-related keyword (which is the primary method we use in this paper to calculate GenAI adoption). We then compare this job posting-based measure to (1) the occupation-level adoption rates reported by Bick et al. (2024) and (2) the Eloundou et al. (2023) predicted-exposure metric, both as presented in Bick et al.’s Figure 8.

Figure C2 below reports the results. Although the keyword-based posting shares are lower in level, they are strongly correlated with Bick et al. (2024) adoption rates ($\rho = 0.720$), as shown in Figure C2.a. These patterns suggest that job postings capture substantial cross-occupation variation in GenAI adoption while possibly understating adoption levels overall. This hints at the possibility that some GenAI-adopting roles do not explicitly reference GenAI in postings and our regression estimates built on these data are therefore likely conservative. When benchmarking against the Eloundou et al. (2023) predicted-exposure metric, the correlation between the two measures is also strong ($\rho = 0.724$), as shown in Figure C2.b.

**Figure C1.** Illustrative job postings to identify *GenAI roles*.

GenAI - UX Designer

At Lilly, we unite caring with discovery to make life better for people around the world. We are a global healthcare leader headquartered in Indianapolis, Indiana. Our employees around the world work to discover and bring life-changing medicines to those who need them, improve the understanding and management of disease, and give back to our communities through philanthropy and volunteerism. We give our best effort to our work, and we put people first. We're looking for people who are determined to make life better for people around the world.

Generative AI - Proactive, Predictive, Powerful!

Lilly is actively looking for an experienced UX Designer who will embrace the power of GenAI to innovate and enhance the creation of design concepts. Are you passionate about crafting the future of UX Design through Generative AI? If so, bring YOUR skills and talents to Lilly where you'll have the chance to create an impact on the lives of patients!

What You'll Be Doing
As a GenAI UX Designer, you will have the technical accountability for the design and creation of new digital assets. You will work closely with creative agencies, marketing teams, and other digital developers to create optimal user experiences within digital tactics. You will create a sound approach for the evaluation, creation, and delivery of digital assets to promote products and provide medical information.

How You'll Succeed
Apply your front-end development skills to create highly functioning digital assets.
Provide insight and mentorship on the feasibility of newly created digital assets.
Design prototypes to guide development.
Create engaging and interactive user interfaces that follow the Lilly Design System.
Own and maintain code driving good software project management and development practices.
Collaborate with business partners to create a cohesive and visually appealing user interface.
Provide leadership and guidance on UI/UX standard methodologies for Lilly developers and our external partners
Conduct research to identify emerging trends and technologies in GenAI and UX.

What You Should Bring
Experience in designing AI driven user interfaces and experiences
Validated knowledge and experience with implementing user experience design principles.
Solid grasp of React, AngularJS.
Experience working on projects involving the creation, use or implementation of design systems.
Strong conceptualization and visual communication ability.
Superb written and verbal communication skills.
Strong collaboration skills.
Shown organizational skills.
Understanding of cultural design considerations for global consumption of information.
Portfolio of UX work showcasing design process and execution and demonstrates how your contributions informed or improved the final product.

**Figure C1(a).** Illustrative Job Posting 1

LexisNexis Risk Solutions is the essential partner in the assessment of risk. Within our Business Services vertical, we offer a multitude of solutions focused on helping businesses of all sizes drive higher revenue growth, maximize operational efficiencies, and improve customer experience.

About the Role

As our Innovation Integration Specialist, you will lead the integration of cutting-edge technologies, such as AI, machine learning, automation, and other emerging innovations, into our operational processes.

Your expertise in practical technology implementation, advanced analytics, and requirement writing will bridge the gap between technical capabilities and business objectives.

This dynamic role combines strategic planning, hands-on implementation, and continuous learning and advocacy for innovative technologies. Each day offers a blend of collaborative work, technical assessments, and opportunities to drive significant improvements in operational efficiency and effectiveness.

Responsibilities

Serving as a champion for innovation, educating and inspiring others about the potential benefits while critically evaluating existing and new technology solutions for suitability across various use cases.

Collaborating with business stakeholders to write requirements and bridge the gap between technical needs and business objectives.

Implementing innovative tools and technologies, including AI, automation, and more, within the company's operational framework.

Maintaining an up-to-date understanding of the technology landscape to inform decision-making and strategy development.

Ensuring technology implementations comply with company restrictions and regulatory environments.

Utilizing advanced analytics to derive actionable insights and improve decision-making processes.

Extracting and merging data from multiple databases, generate analytics, and create comprehensive reports on KPIs.

Requirements

Demonstrate proficiency in advanced analytics, data management, and deriving actionable insights.

Hands-on experience in developing and automating test scripts, bots, and workflows to streamline operations and boost efficiency, combined with prior expertise in creating AI-driven chatbots, virtual assistants, and utilizing AI tools such as Copilot 365 and ChatGPT.

Experience working with stakeholders to collect and analyze data, develop recommendations, and deliver data-driven solutions.

Demonstrate thinker with excellent communication and presentation skills, and the ability to convey technical concepts in a understandable manner.

Have experience with innovative technology projects and tools such as AWS ecosystem, SQL, and Python.

Experience with Salesforce, Planview PPM Pro, JIRA, DocuSign

**Figure C1(b).** Illustrative Job Posting 2

**Figure C2.** Scatterplot between Lightcast GenAI adoption measure and (1) Figure C2(a): Bick et al. (2024) adoption metric, and (2) Figure C2(b): Eloundou et al. (2023) predicted exposure metric.

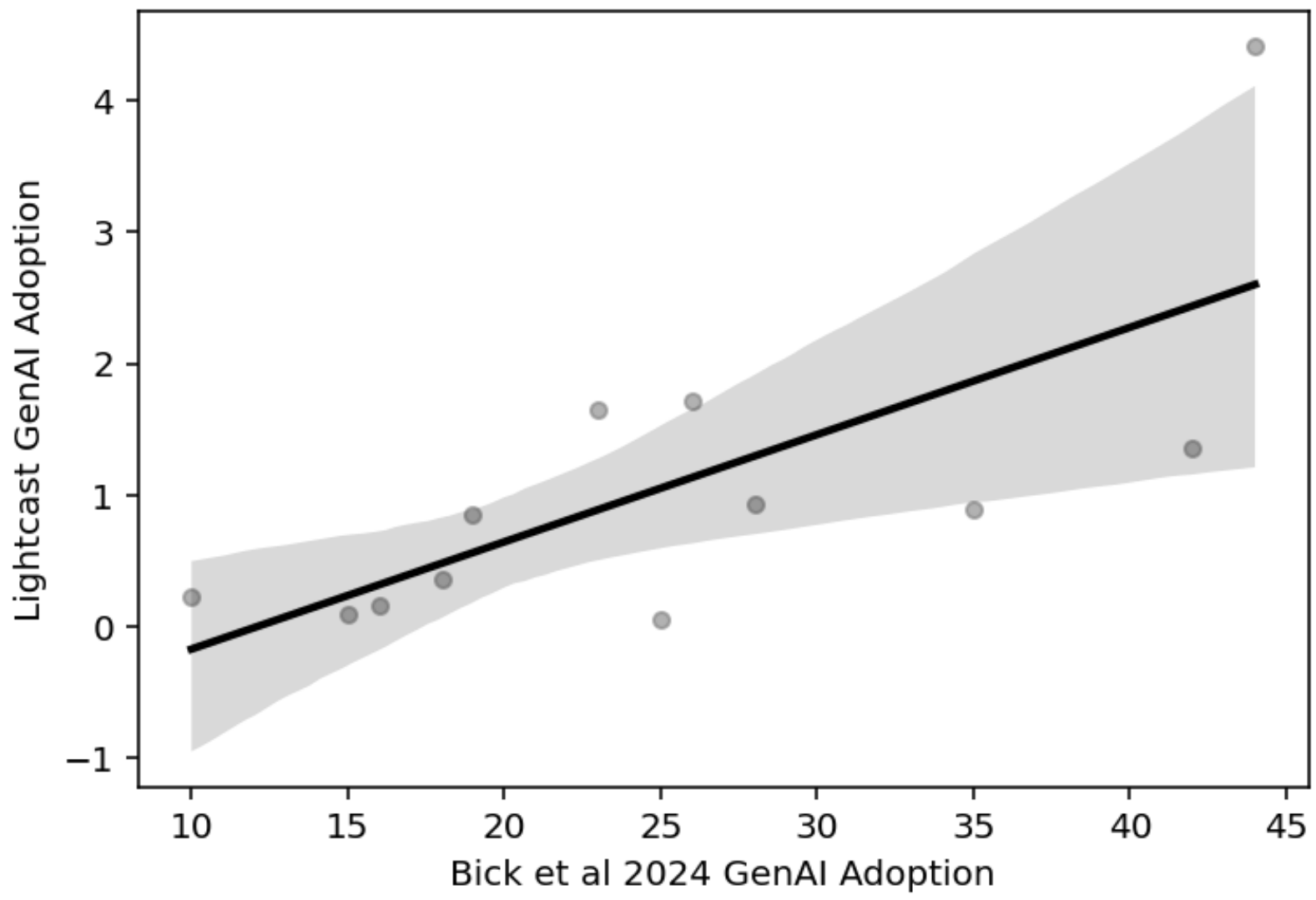

**Figure C2(a).** Bick et al. (2024) GenAI adoption vs Lightcast GenAI adoption

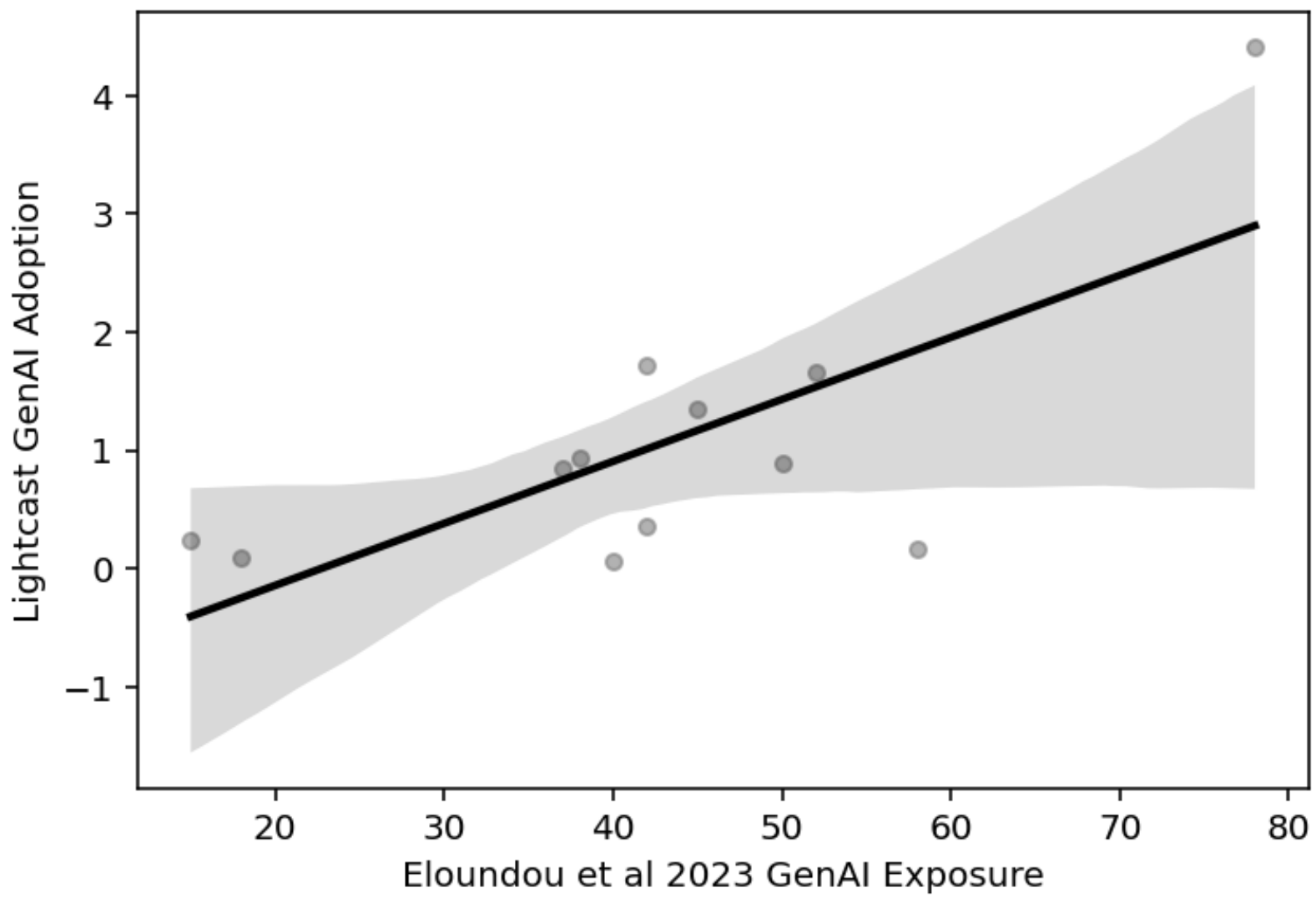

**Figure C2(b).** Eloundou et al. (2023) GenAI predicted exposure vs Lightcast GenAI adoption

**Appendix D: Time Trends**

We verify our main results by plotting time trends for the treated and counterfactual groups separately for changes in Social Skills Intensity and Other Skills Count (Inverse Hyperbolic Sine Transformation). These are plotted in Figures D1 and D2 below.

**Figure D1.** Time trends for Social Skills Intensity plotted separately for the Treated (GenAI roles) and Counterfactual Group (Other Roles).

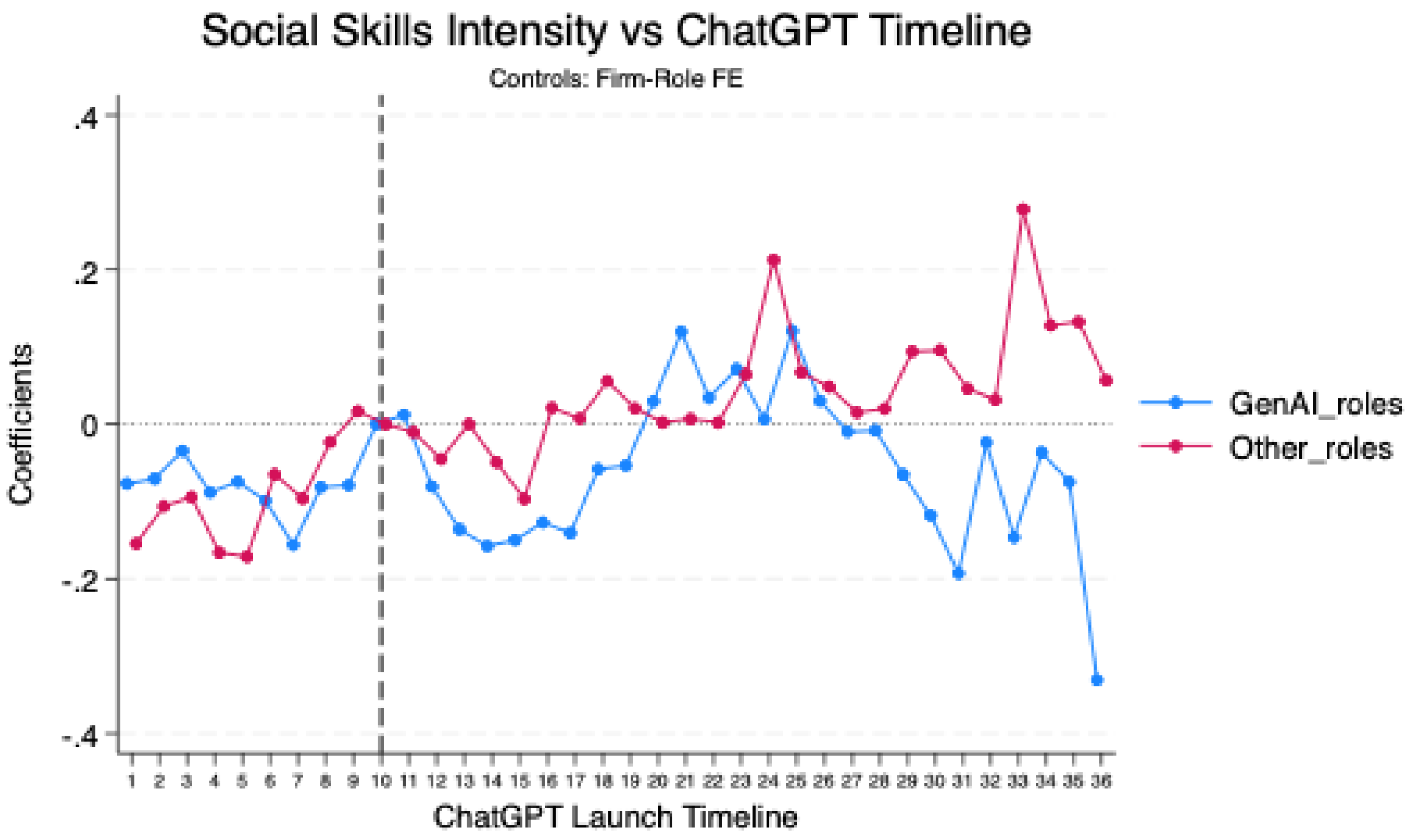

**Figure D2.** Time trends for Other Skills Count plotted separately for the Treated (GenAI roles) and Counterfactual Group (Other Roles).

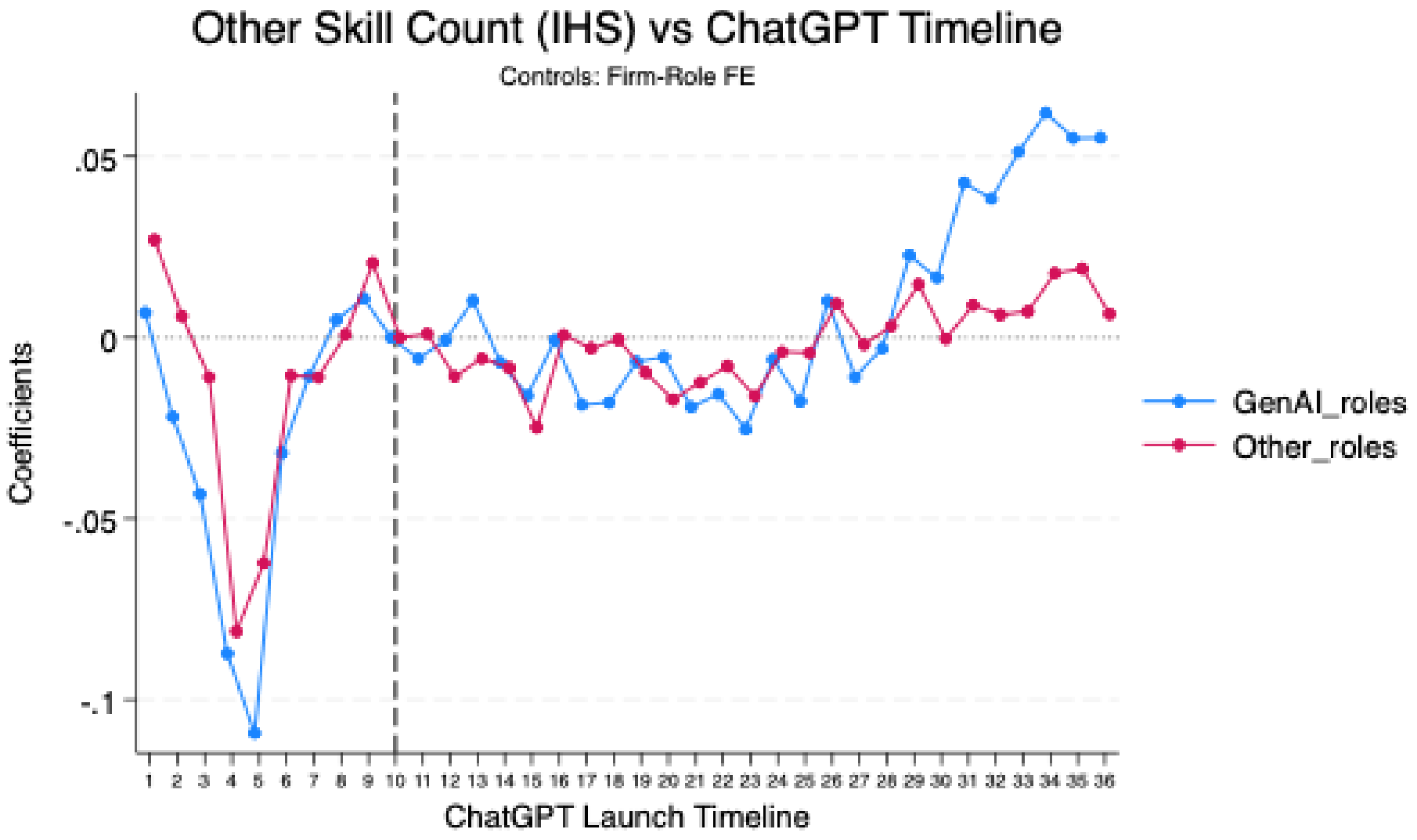

## Appendix E: Robustness Tests

We check whether our results are robust to (1) excluding GenAI-related skills from the skill taxonomy, (2) dropping firms in GenAI-producing or consulting sectors, and (3) dropping technical roles. These results are reported in Tables E1-E3, respectively.

**Table E1.** Main Results: Robustness after excluding GenAI-related skills from skill counts.

| | (1) Cognitive Skills Intensity | (2) Social Skills Intensity | (3) Cognitive Skills Count | (4) Social Skills Count | (5) Other Skills Count |
|---|---|---|---|---|---|
| GenAI Role × Post ChatGPT Launch | -0.048 | -0.110 | -0.005 | -0.008 | 0.024 |
| | (0.121) | (0.040) | (0.015) | (0.009) | (0.011) |
| | [0.693] | [0.006] | [0.723] | [0.391] | [0.036] |
| Firm-Role FE | Y | Y | Y | Y | Y |
| Month FE | Y | Y | Y | Y | Y |
| Observations | 1,709,731 | 1,709,731 | 1,709,731 | 1,709,731 | 1,709,731 |
| Num firms | 595 | 595 | 595 | 595 | 595 |
| Num roles | 41,852 | 41,852 | 41,852 | 41,852 | 41,852 |
| Adj R-square | 0.652 | 0.663 | 0.707 | 0.695 | 0.780 |
| Mean DV | 2.939 | 3.273 | 0.455 | 0.498 | 3.309 |

Note: Standard errors, clustered at the firm and role level, in parentheses. p-values are reported in square brackets below the standard errors.

**Table E2.** Main Results: Robustness after excluding NAICS 51,54 firms (i.e., firms in GenAI-producing or consulting sectors).

| | (1) Cognitive Skills Intensity | (2) Social Skills Intensity | (3) Cognitive Skills Count | (4) Social Skills Count | (5) Other Skills Count |
|---|---|---|---|---|---|
| GenAI Role × Post ChatGPT Launch | -0.071 | -0.106 | -0.010 | -0.009 | 0.020 |
| | (0.143) | (0.038) | (0.018) | (0.010) | (0.012) |
| | [0.619] | [0.005] | [0.595] | [0.370] | [0.103] |
| Firm-Role FE | Y | Y | Y | Y | Y |
| Month FE | Y | Y | Y | Y | Y |
| Observations | 1,409,842 | 1,409,842 | 1,409,842 | 1,409,842 | 1,409,842 |
| Num firms | 443 | 443 | 443 | 443 | 443 |
| Num roles | 39,795 | 39,795 | 39,795 | 39,795 | 39,795 |
| Adj R-square | 0.662 | 0.671 | 0.722 | 0.708 | 0.792 |
| Mean DV | 2.900 | 3.312 | 0.442 | 0.496 | 3.273 |

Note: Standard errors, clustered at the firm and role level, in parentheses. p-values are reported in square brackets below the standard errors.

**Table E3.** Main Results: Robustness for non-technical roles only (i.e., roles outside 11-Managerial Occupations and 15-Computer and Mathematical Occupations).

| | (1) Cognitive Skills Intensity | (2) Social Skills Intensity | (3) Cognitive Skills Count | (4) Social Skills Count | (5) Other Skills Count |
|---|---|---|---|---|---|
| GenAI Role × Post ChatGPT Launch | -0.005 | -0.132 | 0.003 | -0.007 | 0.024 |
| | (0.111) | (0.039) | (0.011) | (0.010) | (0.008) |
| | [0.963] | [0.001] | [0.757] | [0.481] | [0.005] |
| Firm-Role FE | Y | Y | Y | Y | Y |
| Month FE | Y | Y | Y | Y | Y |
| Observations | 996,995 | 996,995 | 996,995 | 996,995 | 996,995 |
| Num firms | 595 | 595 | 595 | 595 | 595 |
| Num roles | 27,057 | 27,057 | 27,057 | 27,057 | 27,057 |
| Adj R-square | 0.663 | 0.688 | 0.729 | 0.729 | 0.801 |
| Mean DV | 2.689 | 3.317 | 0.381 | 0.457 | 3.144 |

Note: Standard errors, clustered at the firm and role level, in parentheses. p-values are reported in square brackets below the standard errors.

## Appendix F: Additional Supplementary Analyses

To examine which specific skills contribute to the skill reweighting effect that drives the observed decline in social skills intensity, we further study the "other skills" category (i.e., all skills except social and cognitive) by decomposing them into skill sub-groups. The Lightcast proprietary skill taxonomy identifies granular skills in each job posting and also maps such individual skills into aggregate skill subcategories and skill categories. Given our objective of studying skill sub-groups, we use such higher-level categories rather than the individual skills.

We aggregate Lightcast *skill categories* into seven groups: customer service, manufacturing operations, research and development, information technology, functional administrative skills, general administrative skills, and other industry-specific skills. Functional administrative skills comprise finance, human resources, and legal skills. We draw on Guadalupe et al. (2014) to identify primary and administrative functions within organizations. We use these to classify skill categories into five functional sub-groups (customer service, manufacturing operations, research and development, information technology, and functional administrative skills). We then group the remaining general skill categories, related to business, administration, and policy, into general administrative skills and classify the remaining specialized skill categories as other industry-specific skills. Table F1 reports the resulting mapping. Table F2 provides examples of the top five skills within each mapped skill group.

Further, to examine whether the skill reweighting effect is predominantly driven by skill deepening of functional administrative skills within finance, human resources, or legal occupations, or skill extension within other occupations, we partition our sample into these two occupation groups manually, based on their mapped six-digit O*NET-SOC occupation. Table F3 provides the SOC codes designated as finance, human resources, and legal occupations.

**Table F1.** Lightcast skill category mapped onto the analyzed skill groups.

| Lightcast skill category name | Mapped skill group |
|---|---|
| Customer and Client Support | Customer Service |
| Hospitality and Food Services | Customer Service |
| Marketing and Public Relations | Customer Service |
| Property and Real Estate | Customer Service |
| Sales | Customer Service |
| Finance | Functional Administrative |
| Human Resources | Functional Administrative |
| Law, Regulation, and Compliance | Functional Administrative |
| Administration | General Administrative |
| Business | General Administrative |
| Economics, Policy, and Social Studies | General Administrative |
| Architecture and Construction | Manufacturing Operations |
| Energy and Utilities | Manufacturing Operations |
| Environment | Manufacturing Operations |
| Maintenance, Repair, and Facility Services | Manufacturing Operations |
| Manufacturing and Production | Manufacturing Operations |
| Transportation, Supply Chain, and Logistics | Manufacturing Operations |
| Science and Research | Research and Development |
| Analysis | Technological |
| Design | Technological |
| Engineering | Technological |
| Information Technology | Technological |
| Agriculture, Horticulture, and Landscaping | Other Industry-specific |
| Education and Training | Other Industry-specific |
| Health Care | Other Industry-specific |
| Media and Communications | Other Industry-specific |
| Performing Arts, Sports, and Recreation | Other Industry-specific |
| Personal Care and Services | Other Industry-specific |
| Physical and Inherent Abilities | Other Industry-specific |
| Public Safety and National Security | Other Industry-specific |
| Social and Human Services | Other Industry-specific |

**Table F2.** Top five individual skills within each analyzed skill group.

| Individual skill name | Mapped skill group |
|---|---|
| Merchandising | Customer Service |
| Marketing | Customer Service |
| Selling Techniques | Customer Service |
| Cash Register | Customer Service |
| Product Knowledge | Customer Service |
| Auditing | Functional Administrative |
| Finance | Functional Administrative |
| Accounting | Functional Administrative |
| Regulatory Compliance | Functional Administrative |
| Cash Handling | Functional Administrative |
| Project Management | General Administrative |
| Workflow Management | General Administrative |
| Process Improvement | General Administrative |
| Continuous Improvement Process | General Administrative |
| Customer Relationship Management | General Administrative |
| Warehousing | Manufacturing Operations |
| New Product Development | Manufacturing Operations |
| Inventory Management | Manufacturing Operations |
| Supply Chain | Manufacturing Operations |
| Palletizing | Manufacturing Operations |
| Biology | Research and Development |
| Clinical Trials | Research and Development |
| Biotechnology | Research and Development |
| Chemistry | Research and Development |
| Life Sciences | Research and Development |
| Computer Science | Technological |
| Agile Methodology | Technological |
| Automation | Technological |
| Data Analysis | Technological |
| SQL (Programming Language) | Technological |
| Pharmaceuticals | Other Industry-specific |
| Healthcare Industry Knowledge | Other Industry-specific |
| Nursing | Other Industry-specific |
| Safety Training | Other Industry-specific |
| Medicare | Other Industry-specific |

**Table F3.** O*NET SOCs designated as finance, human resources, and legal occupations

| SOC code | SOC code description | Mapped function |
|---|---|---|
| 11-3031.00 | Financial Managers | Finance |
| 11-3031.01 | Treasurers and Controllers | Finance |
| 13-2051.00 | Financial and Investment Analysts | Finance |
| 13-2052.00 | Personal Financial Advisors | Finance |
| 13-2054.00 | Financial Risk Specialists | Finance |
| 13-2061.00 | Financial Examiners | Finance |
| 13-2099.00 | Financial Specialists, All Other | Finance |
| 13-2099.01 | Financial Quantitative Analysts | Finance |
| 13-2099.04 | Fraud Examiners, Investigators and Analysts | Finance |
| 43-3099.00 | Financial Clerks, All Other | Finance |
| 11-3121.00 | Human Resources Managers | Human Resources |
| 11-3131.00 | Training and Development Managers | Human Resources |
| 13-1071.00 | Human Resources Specialists | Human Resources |
| 13-1151.00 | Training and Development Specialists | Human Resources |
| 43-4161.00 | Human Resources Assistants, Except Payroll and Timekeeping | Human Resources |
| 11-1031.00 | Legislators | Legal |
| 23-1011.00 | Lawyers | Legal |
| 23-1012.00 | Judicial Law Clerks | Legal |
| 23-1021.00 | Administrative Law Judges, Adjudicators, and Hearing Officers | Legal |
| 23-1022.00 | Arbitrators, Mediators, and Conciliators | Legal |
| 23-1023.00 | Judges, Magistrate Judges, and Magistrates | Legal |
| 23-2011.00 | Paralegals and Legal Assistants | Legal |
| 23-2093.00 | Title Examiners, Abstractors, and Searchers | Legal |
| 23-2099.00 | Legal Support Workers, All Other | Legal |